\documentclass[a4paper,11pt]{article}
\usepackage[hhmmss]{datetime}
\usepackage{jheppub}
\hypersetup{colorlinks=true,linkcolor=magenta,anchorcolor=green,citecolor=cyan,filecolor=black,menucolor=black,urlcolor=brown}

\usepackage[table]{xcolor}
\usepackage{slashbox}
\allowdisplaybreaks
\usepackage{graphicx}
\usepackage{epstopdf}
\usepackage[utf8]{inputenc} 
\usepackage{tikz}
\usepackage{tikz-cd}
\usetikzlibrary{shapes.arrows,calc,decorations.pathreplacing,decorations.markings,bending,arrows,positioning}
\tikzset{
    >=stealth',
    punkt/.style={
           rectangle,
           rounded corners,
           draw=black, very thick,
           text width=6.5em,
           minimum height=2em,
           text centered},
    pil/.style={
           ->,
           thick,
           shorten <=2pt,
           shorten >=2pt,}
}

\newcommand{\be}{\begin{equation}}
\newcommand{\ee}{\end{equation}}
\newcommand{\bea}{\begin{eqnarray}}
\newcommand{\eea}{\end{eqnarray}}
\newcommand{\nn}{\nonumber}

\newcommand{\bdm}{\begin{displaymath}}
\newcommand{\edm}{\end{displaymath}}

\usepackage[force]{feynmp-auto}

\title{The Schwarzschild-Tangherlini metric from scattering amplitudes
in various dimensions}

\author[a]{Stavros Mougiakakos}
\author[a,b,c]{\textrm{and}  Pierre Vanhove}

\affiliation[a]{Institut de Physique Th\'eorique,\\
  Universit\'e Paris Saclay, CNRS, F-91191 Gif-sur-Yvette, France}
\affiliation[b]{National Research University Higher School of Economics, Russian Federation}
\affiliation[c]{Theoretical Physics Department, CERN, 1211 Geneva 23, Switzerland}
\preprint{\vbox{\hbox{\hphantom{XXXX}IPhT-t20/053}\hbox{CERN-TH-2020-168}}}

\abstract{
  We derive the static Schwarzschild-Tangherlini metric by extracting
  the classical contributions from the
  multi-loop vertex functions of a graviton emitted from a massive
  scalar field.    At each loop orders the classical contribution is
  proportional to a unique master integral given by the massless
  sunset integral. By computing the scattering amplitudes up to
  three-loop order in general dimension, we explicitly derive the expansion of the metric
  up to the fourth post-Minkowskian order $O(G_N^4)$  in four, five
  and six dimensions.  There are  ultraviolet divergences that are cancelled  with the
  introduction of higher-derivative non-minimal couplings.  The 
  standard  Schwarzschild-Tangherlini  is recovered by absorbing their
  effects  by  an appropriate coordinate transformation induced from the
  de Donder gauge condition. }

\begin{document}

\today

\maketitle

%\flushbottom
\newpage

\section{Introduction} 

General relativity is a theory for the action of gravity in space and
time. The dynamics of the gravitational field is constrained by the
Einstein's classical field equations. They are tensorial non-linear
equations, because of the self-interaction of the gravitational field,
notoriously difficult to solve. It is therefore important to develop
efficient methods for studying gravity in various regimes.

General relativity can be embedded in quantum theory where the gravitational
force results from the exchange of  a quantized massless spin-2 graviton
field~\cite{tHooft:1974toh,Veltman:1975vx,DeWitt:1967yk,DeWitt:1967ub,DeWitt:1967uc}. One
can 
then consider the Einstein-Hilbert term as the first term of a low-energy
effective action containing an infinite number of higher derivative operators~\cite{Donoghue:1994dn}.

The classical limit $\hbar\to0$ has been studied by Duff
in~\cite{Duff:1973zz} where he showed how to reproduce the classical Schwarzschild 
metric in four dimensions from quantum tree graphs up to
the second order $O(G_N^2)$ in Newton's constant.

The relation between the quantum theory of gravity and the classical
Einstein's theory of general relativity has received a new
interpretation with the  understanding~\cite{Iwasaki:1971vb,BjerrumBohr:2002ks,Holstein:2004dn,Donoghue:1996mt,Bjerrum-Bohr:2018xdl,Kosower:2018adc} that  an appropriate (and
subtle) $\hbar\to0$ limit of quantum multi-loop
scattering gravitational amplitudes lead to higher $G_N$-order
classical gravity contributions. Considering the importance of such
approach for the evaluation of  the post-Minkowskian  expansion for
the gravitational two-body scattering~\cite{Cheung:2018wkq, Bern:2019nnu,Bern:2019crd,Chung:2019duq,Kalin:2019rwq,Cheung:2020gyp,1821624}, we use the procedure given
in~\cite{Bjerrum-Bohr:2018xdl} for extracting the classical
contributions from the multi-loop vertex function of a graviton emission from a
massive scalar field to recover the Schwarzschild-Tangherlini  metric in
various dimensions.
The scattering amplitude  approach works in general
dimensions~\cite{Collado:2018isu,KoemansCollado:2019ggb,Cristofoli:2020uzm,Jakobsen:2020ksu}
and gives the opportunity to explore general relativity in higher-dimensions~\cite{Emparan:2008eg,Emparan:2013moa}.  At tree-level and one-loop our results agree
with the general dimension results
in~\cite{Collado:2018isu,Jakobsen:2020ksu}.  We show how to
reconstruct the metric up to the fourth order $O(G_N^4)$ in Newton's
constant by evaluating the scattering amplitudes up to three-loop orders.

Using the procedure designed in~\cite{Bjerrum-Bohr:2018xdl} we argue,
in section~\ref{sec:class-contr}, that the classical contribution at
$l$-loop order is given by the two-point $l$-loop massless sunset
graphs. We verify this explicitly
evaluating the classical limit of the quantum scattering amplitudes  up to
three-loop order.

The scattering amplitudes develop ultraviolet divergences.   In section~\ref{sec:nonmin}, we show how 
to recover the finite static Schwarzschild-Tangherlini metric by the addition of non-minimal
couplings given schematically by  (see~\eqref{e:Sctn} 
for  a precise expression)
\begin{equation}
\delta^{(n)}S^{\rm ct.} \sim (G_N m)^{2n\over d-2}  \int d^{d+1}x \sqrt{-g}   \,
\nabla^{2(n-1)} \mathcal R_{\mu\nu}  \partial^\mu \phi \partial^\nu\phi\,.
\end{equation}
In four dimensions the non-minimal couplings $\delta^{(1)}S^{\rm ct.}$ have been introduced
in~\cite{Goldberger:2004jt} for the analysis up to the third
post-Minkowskian order in the context of the world-line formalism. The
relation between the world-line formalism and the amplitude
approach is detailed in~\cite{1821624}.
Higher-derivative couplings with $n\geq2$  would be needed in four
dimensions from the fifth post-Minkowskian order, but they appear at
lowest order in higher dimensions. 
Indeed, we show that in five
dimensions one needs to consider  higher dimensional of non-minimal
couplings $\delta^{(2)}S^{\rm ct.}$ at the third post-Minkowskian
order and $\delta^{(3)}S^{\rm ct.}$ at the fourth post-Minkowskian.  Interestingly, the metric
components are finite in space-time dimensions greater or equal to six, although the stress-tensor
develops  ultraviolet divergences from one-loop order in odd dimensions and
from two-loop order in even dimensions. These
divergences are cancelled by the non-minimal couplings
$\delta^{(n)}S^{\rm ct.}$.
Actually, we expect
that an all order computation in perturbation will require an
infinite set of such non-minimal couplings.

We show that the effects of the non-minimal couplings can be
reabsorbed by a coordinate transformation, and they do not affect the
Schwarzschild-Tangherlini space-time geometry.  Since we work in
the fixed gauge de Donder gauge, we give the coordinate transformation
for extracting the classical space-time metric from the scattering amplitudes in
that gauge. Although general
relativity is coordinate system invariant, our analysis shows that
there is a preferred coordinate system when extracting the classical
geometry from scattering amplitudes in the de Donder gauge. The
lowest-order $n=1$ non-minimal couplings have been shown to arise from
the gauge fixing in~\cite{BjerrumBohr:2006mz,Jakobsen:2020ksu,1821624}.  We will not address the question of the gauge dependence,
but we remark that the choice of coordinate system (or gauge) can be
critical for finding solution to Einstein's
equations~\cite{Fromholz:2013hka}.

Since ``black hole formation is a
  robust prediction of the general theory of
  relativity''~\cite{nobelpenrose}, it is satisfying to be able to
  embed such classical solutions in the new understanding of the
  relation between general relativity and the quantum theory of gravity. 

  The paper is organised as follows. In section~\ref{sec:schw} we setup
  the connection between the perturbation expansion vertex function
  for the emission a graviton from a massive scalar field and the
  post-Minkowskian expansion of the static metric in $d+1$ dimensions.
  In section~\ref{sec:class-contr} we
show that the classical contribution from the
multi-loop amplitudes is given by the massless sunset multi-loop
integrals in $d$ dimensions. In
section~\ref{sec:mast-integr-class} we evaluate the master integrals. In section~\ref{sec:class-metr-pert} we
derive the metric component up to the order $O(G_N^4)$ by computing
the relevant amplitudes up to three-loop order in $d+1$ dimensions.
In section~\ref{sec:nonmin} we compute  the non-minimal couplings
required for cancelling the ultraviolet divergences in the amplitude computation.
In
section~\ref{sec:deDonder} we solve the Einstein's equations in
four ($d=3$), five ($d=4$) and six ($d=5$) dimensions in the de Donder gauge, and we show
in  section~\ref{sec:matching-amplitude} how these results match the results derived from the amplitude computations.
In
section~\ref{sec:discussion} we give 
an interpretation of the results in this paper. The appendix~\ref{sec:FT}
contains  formul\ae{} for the Fourier transforms used in the text,
and appendix~\ref{sec:vertices} the vertices for the scattering
amplitude computations.

\section{The Schwarzschild-Tangherlini metric from scalar field amplitudes}
\label{sec:schw}
The Schwarzschild metric is obtained by the gravitational scattering
of a scalar field of mass $m$

\begin{equation}
\mathcal{S}=\int d^{d+1}x \sqrt{-g}\left({R\over 16\pi G_N}+
\frac{1}{2} g^{\mu\nu}\partial_{\mu}\phi \partial_{\nu}\phi-\frac{1}{2}m^2\phi^2\right)\,.
\end{equation}
For further reference Newton's constant has length dimensions
$[G_N]=(length)^{d-1}$, the scalar field has dimension
$[\phi]=(length)^{1-d}$ and the mass $[m]=(length)^{-1}$. We work with
 the mostly negative signature   $(+,-,\cdots,-)$ metric.

The graviton emission from a scalar particle of mass $p_1^2=p_2^2=m^2$
is  given by the three-point vertex function 
\begin{equation}\label{e:M3pt}
\mathcal M_3(p_1,q)  =\qquad
\begin{gathered}
\begin{fmffile}{gravemission}
    \begin{fmfgraph*}(100,100)
\fmfstraight
\fmfleftn{i}{2}
\fmfrightn{o}{1}
\fmftop{v3}
\fmfbottom{v4}
\fmfrpolyn{smooth,filled=30}{G}{3}
\fmf{fermion,tension=2.5,label=$p_1$}{i1,G1}
\fmf{fermion,tension=2.5,label=$p_2$}{G2,i2}
\fmf{dbl_wiggly,tension=.5,label=${q}\qquad$}{G3,o1}
\end{fmfgraph*}
\end{fmffile}
\end{gathered}\quad .
\end{equation}
At each loop order we  extract the $l$-loop contribution to the
transition density   of the stress-energy tensor
$\langle T_{\mu\nu}(q^2)\rangle=\sum_{l\geq0} \langle T^{(l)}_{\mu\nu}(q^2)\rangle$
\begin{equation}\label{e:MtoT}
  i\mathcal M^{ (l )}_3(p_1,q)  =-{i\sqrt{32\pi G_N}\over2}
\langle  T^{(l)\, \mu\nu}(q^2)  \rangle\epsilon_{\mu\nu}
\end{equation}
where $\epsilon^{\mu\nu}$ is the polarisation of the graviton with
momentum $q=p_1-p_2$ is the   momentum transfer.

The  scattering amplitude computation is not done in the harmonic  gauge
coordinates $g^{\mu\nu}\Gamma^\lambda_{\mu\nu}(g)=0$  but in  the \textit{de
  Donder gauge}
coordinate system~\cite{Veltman:1975vx,Goldberger:2004jt,Cheung:2020gyp,Collado:2018isu,Jakobsen:2020ksu}
\begin{equation}\label{e:deDonderGauge}
\eta^{\mu\nu}\Gamma^\lambda_{\mu\nu}(g)= \eta^{\mu\nu}g^{\lambda\rho}\left({\partial g_{\rho\mu}\over \partial x^\nu}+ {\partial g_{\rho\nu}\over \partial x^\mu}-{\partial g_{\mu\nu}\over \partial x^\rho}\right)=0\,,
\end{equation}
the metric  perturbations $g_{\mu\nu}=\eta_{\mu\nu} +\sum_{n\geq1}
h^{(n)}_{\mu\nu}$ satisfy\footnote{The   harmonic gauge linearized
  at the first order in perturbation gives~(\ref{e:hndD}) with
  $n=1$. The higher-order expansions of the harmonic
gauge differ from these conditions.}
\begin{equation}\label{e:hndD}
    {\partial\over
  \partial x^{\lambda}} h^{\lambda (n)}_{\nu}-\frac{1}{2} {\partial\over
  \partial x^{\nu}} h^{(n)}         =0\,.
\end{equation}
 The de Donder gauge relation between the
metric perturbation and the stress-energy tensor reads
\begin{equation}\label{e:TtohAmplitudedeDonder}
 h^{(l+1)}_{\mu\nu}(\vec x) = -16\pi G_N\int {d^d{\vec q}\over(2\pi)^d} e^{i\vec q\cdot \vec x} {1\over
 \vec q^2} \left( \langle T_{\mu\nu}^{(l)}\rangle^{\rm
   class.}(q^2)-\frac{1}{d-1}\eta_{\mu\nu}\langle T^{(l)}\rangle^{\rm class.}(q^2)\right)\,.
\end{equation}
In this relation enters the classical contribution at $l$ loop order $\langle T^{(l)}_{\mu\nu}\rangle^{\rm class.}(q^2) $ defined
by the classical limit of the quantum scattering
amplitude~\cite{Holstein:2004dn,Bjerrum-Bohr:2018xdl,Kosower:2018adc}.
From now, we are dropping the super-script class and just use the
notation $\langle T^{(l)}_{\mu\nu}\rangle(q^2) $   for the classical contribution.

%------------------------------------------------------------------------
\subsection{The classical contribution of the amplitude}
\label{sec:class-contr}

In this section we derive the generic form
of the classical contribution of the gravity amplitudes~(\ref{e:M3pt}) in the static limit where $q=(0,\vec
q)$ and $\vec q^2\ll m^2$. The classical limit is obtained by taking
$\hbar\to0$ with the momentum transfer $q/\hbar$  held fixed~\cite{Kosower:2018adc}.

At the $l$-loop order  we have to consider the
graphs 

\begin{equation}\label{e:M3quantum}
\mathcal M^{(l)}_3(p_1,q)  =
\begin{gathered}
\begin{fmffile}{gravemissionlloop}
    \begin{fmfgraph*}(100,100)
\fmfstraight
\fmfleftn{i}{6}
\fmfrightn{o}{1}
\fmfrpolyn{smooth,label={tree},filled=30}{G}{5}
\fmf{dbl_wiggly}{G1,i2}
\fmf{dbl_wiggly}{G2,i3}
\fmf{dbl_wiggly}{G3,i4}
\fmf{dbl_wiggly}{G4,i5}
\fmf{dbl_wiggly,tension=3,label={q}}{G5,o1}
\fmf{plain,tension=2.5}{i1,i2,i3,i4,i5,i6}
%\fmf{fermion,tension=2.5,label=$p_2$}{v2,i2}
\end{fmfgraph*}
\end{fmffile}
\end{gathered}\,,
\end{equation}
The classical contribution emerges as
a particular $\hbar \to 0$ limit of the amplitude
in~\cite{Iwasaki:1971vb,Holstein:2004dn,Donoghue:1996mt,Kosower:2018adc,Bjerrum-Bohr:2018xdl}.
The  classical limit results in cutting the massive lines, 
projecting on the contribution from  localised sources at  different
positions in space~\cite{PlanteThesis,Galusha,Bjerrum-Bohr:2018xdl},
pictorially represented by shaded blobs

\begin{equation}\label{e:M3classical}
\mathcal M^{(l)~\rm class.}_3(p_1,q)  =\qquad
\begin{gathered}
\begin{fmffile}{gravemissionlloopclassical}
    \begin{fmfgraph*}(100,100)
\fmfstraight
\fmfleftn{i}{6}
\fmfrightn{o}{1}
\fmfrpolyn{smooth,label={tree},filled=30}{G}{5}
\fmf{dbl_wiggly}{G1,i2}
\fmf{dbl_wiggly}{G2,i3}
\fmf{dbl_wiggly}{G3,i4}
\fmf{dbl_wiggly}{G4,i5}
\fmf{dbl_wiggly,tension=3,label={q}}{G5,o1}
\fmfblob{.5cm}{i2}
\fmfblob{.5cm}{i3}
\fmfblob{.5cm}{i4}
\fmfblob{.5cm}{i5}
\end{fmfgraph*}
\end{fmffile}
\end{gathered}\Bigg|_{\textrm{leading}~q^2}\,,
\end{equation}

In this process one keeps only the leading $q^2$ contribution from the
multi-graviton tree-level amplitudes.
The quantum tree-level graphs that were considered
in~\cite{Duff:1973zz} arise from the classical limit of the scattering
amplitude up to two-loop order. 
In the rest of this section, we derive the
generic features of the classical limit to all orders in perturbation.
We then explicitly evaluate the classical limit up to three-loop order in perturbation.

\medskip

The quantum amplitude in~\eqref{e:M3quantum} is 
 an $l+2$ gravitons amplitude with $l+1$ gravitons attached to the massive scalar
line
\begin{align}
&\mathcal L_{\mu_1\nu_1,\dots,\mu_{l+1}\nu_{l+1}}(p_1,p_2,\ell_1,\dots,\ell_{l+1})  =
\begin{gathered}
\begin{fmffile}{linegraviton}
    \begin{fmfgraph*}(100,100)
\fmfstraight
\fmfleftn{i}{6}
\fmfrightn{o}{6}
\fmf{plain,tension=2.5}{i1,i2,i3,i4,i5,i6}
\fmf{phantom,tension=2.5}{o1,o2,o3,o4,o5,o6}
\fmf{dbl_wiggly}{i2,o2}
\fmf{dbl_wiggly}{i3,o3}
\fmf{dbl_wiggly}{i4,o4}
\fmf{dbl_wiggly}{i5,o5}
\end{fmfgraph*}
\end{fmffile}
\end{gathered}\\
  &={(-i\sqrt{8\pi G_N})^{l+1}\tau_{\mu_1\nu_1}(p_1,p_1-\ell_1)\tau_{\mu_2\nu_2}(p_1\ell_1,p_1-\ell_1-\ell_2)\cdots\tau_{\mu_{l+1}\nu_{l+1}}(p_1-\ell_1-\cdots-\ell_{l+1},p_2)\over \prod_{i=1}^l \left((p_1-\sum_{j=1}^i\ell_j)^2-m^2+i\epsilon\right)}\,,
\end{align}
 with the momentum conservation condition
 $\ell_1+\cdots+\ell_{l+1}=q=p_1-p_2$  and   the vertex for emitting a graviton from a scalar field\footnote{The vertices are given in
  appendix~\ref{sec:vertices}. We have stripped of a factor
  $i\sqrt{8\pi G_N}$ from their normalisation.}
\begin{equation}\label{e:vertex2s1g}
 \tau^{\mu\nu}(p_1,p_2) =p_1^\mu p_2^{\nu}
+p_1^\nu p_2^{\mu} +\frac12 \eta^{\mu\nu}\,(p_1-p_2)^2\,.
\end{equation}
 This line is 
attached to an $l+2$ tree-level graviton amplitude
\begin{equation}
\mathcal M^{\mu_1\nu_1,\dots,\mu_{l+1}\nu_{l+1}}(\ell_1,\dots,\ell_{l+1},q)  =
\begin{gathered}
\begin{fmffile}{gravtreen}
    \begin{fmfgraph*}(100,100)
\fmfstraight
\fmfleftn{i}{6}
\fmfrightn{o}{1}
\fmf{phantom,tension=2.5}{i1,i2,i3,i4,i5,i6}
\fmfrpolyn{smooth,label={tree},filled=30}{G}{5}
\fmf{dbl_wiggly}{G1,i2}
\fmf{dbl_wiggly}{G2,i3}
\fmf{dbl_wiggly}{G3,i4}
\fmf{dbl_wiggly}{G4,i5}
\fmf{dbl_wiggly,tension=3,label={q}}{G5,o1}
\end{fmfgraph*}
\end{fmffile}
\end{gathered}\,.
\end{equation}
We have to sum over all the permutation of the graviton
lines attached to the scalar lines. Because the gravity amplitude is
invariant under  the action of the permutation of the graviton lines
we have
\begin{multline}\label{e:Ml}
 i\mathcal M^{(l)}_3(p_1,q)=\frac{1}{\sqrt{4
     E_1E_2}}\int \prod_{n=1}^l {d^{d+1}\ell_n\over(2\pi)^D}\left( \sum_{\sigma\in\mathfrak S_{l+1}}
  \mathcal
  L_{\mu_1\nu_1,\dots,\mu_{l+1}\nu_{l+1}}(p_1,p_2,\ell_{\sigma(1)},\dots,\ell_{\sigma(l+1)}) \right)
\cr
\times\prod_{i=1}^{l+1} {i\mathcal P^{\mu_i\nu_i,\rho_i\sigma_i}\over \ell_i^2+i\epsilon}\mathcal M_{\rho_1\sigma_1,\dots,\rho_{l+1}\sigma_{l+1}} (\ell_1,\dots,\ell_{l+1},q) 
\end{multline}
where $\mathfrak S_{l+1}$ is the group of permutation of $l+1$ elements.
In the static limit the vertex~(\ref{e:vertex2s1g})
becomes
\begin{equation}
  \tau_{\mu\nu}(p_1,p_1-\ell)\simeq -2 m^2
  \delta^0_\mu \delta^0_\nu\,,
\end{equation}
therefore the scalar line approximates to
\begin{equation}
 \mathcal L(p_1,p_2,\ell_1,\dots,\ell_{l+1})\simeq{ \prod_{i=1}^{l+1} i \sqrt{32\pi
     G_N}m^2\delta^0_{\mu_i}\delta^0_{\nu_i}\over\prod_{i=1}^l\left(
 (p_1-\sum_{j=1}^i \ell_j)^2-m^2+i\epsilon\right)}\,.
\end{equation}
In the static limit $(p_1-L)^2-m^2+i\epsilon= L^2-2 p_1\cdot
L+i\epsilon\simeq L_0^2-\vec L^2-2mL_0+i\epsilon$.
In the limit where  the mass $m$ is large compared to the graviton loop
 momenta $|L|\ll m$ we have 
\begin{multline}
   L_0^2-\vec L^2-2mL_0+i\epsilon= \left(L_0-m-\sqrt{\vec
       L^2+m^2-i\epsilon}\right)
   \left(L_0-m+\sqrt{\vec
       L^2+m^2-i\epsilon}\right)\cr
   \simeq \left(L_0-2m-{\vec L^2\over
       2m}+i\epsilon\right)
   \left(L_0+{\vec L^2\over 2m}-i\epsilon\right)\simeq -2m \left(L_0-i\epsilon\right) \,.
 \end{multline}
 Therefore we have
 \begin{equation}
   \mathcal L(p_1,p_2,\ell_1,\dots,\ell_{l+1})\simeq i\sqrt{32\pi G_N}m^2
   \delta^0_{\mu_{l+1}}\delta^0_{\nu_{l+1}}\prod_{i=1}^{l} {-i 2\sqrt{2\pi
     G_N}m\delta^0_{\mu_i}\delta^0_{\nu_i}\over
 \sum_{j=1}^i \ell^0_j-i\epsilon}\,.
\end{equation}
Using momentum conservation $\ell_1+\cdots+\ell_{l+1}=p_1-p_2$ and
that in the static limit $p_1^0-p_2^0\simeq0$ we have
\begin{equation}
   \mathcal L(p_1,p_2,\ell_1,\dots,\ell_{l+1})\simeq 2m i\epsilon
\prod_{i=1}^{l+1} {-i 2\sqrt{2\pi
     G_N}m\delta^0_{\mu_i}\delta^0_{\nu_i}\over
 \sum_{j=1}^i \ell^0_j-i\epsilon}\,.
\end{equation}
Using the identity\footnote{This was proven in the appendix of~\cite{Levy:1969cr}. We give here an
  alternative proof using recursion. For
$l=1$ we have $\Sigma(2)={1\over x_1(x_1+x_2)}+{1\over x_2(x_1+x_2)}={1\over
  x_1x_2}$. Assuming that~\eqref{e:Sn} is true at the order $l$, then
at the order $l+1$ we have
\begin{equation}
\Sigma(l+1)=  \sum_{\sigma\in\mathfrak S_{l+1}}
  \prod_{i=1}^{l+1}  {1\over \sum_{j=1}^i x_{\sigma(j)}}
  = {1\over x_1+\cdots +x_{l+1}}\sum_{i=1}^{l+1}
  \sum_{\sigma\in\mathfrak S_l} \prod_{i=1}^l {1\over \sum_{j=1}^i
    \hat x_{\sigma(j)}}
\end{equation}
where $\sigma(n+1)=i$ and the $\{\hat x_1,\dots,\hat
x_l\}=\{x_1,\dots,x_{l+1}\}\backslash \{x_i\}$.
By recursion hypothesis we can use the expression for $\Sigma(l)$
\begin{equation}
\Sigma(l+1)
  = {1\over x_1+\cdots +x_{l+1}}\sum_{i=1}^{l+1}
 \prod_{i=1}^{l} {1\over \hat x_i}= {1\over x_1+\cdots +x_{l+1}}\sum_{i=1}^{l+1}
 x_i\prod_{i=1}^{l+1} {1\over x_i}= \prod_{i=1}^{l+1}{1\over x_i}\,.
\end{equation}
}
\begin{equation}\label{e:Sn}
   \sum_{\sigma\in\mathfrak S_{l+1}}
\prod_{i=1}^{l+1}  {1\over \sum_{j=1}^i x_{\sigma(j)}} =
   \prod_{i=1}^{l+1} {1\over 
 x_{i}}\,.
\end{equation}
In the limit $\epsilon\to0$ the expression vanishes unless some
of the $\ell_j^0$ vanish at the same time.
This means that one needs
to pick the residues at $\ell_j^0=i\epsilon$ for  $j=1,\dots,l$ to
have a non vanishing answer.
This implies that the amplitude~(\ref{e:Ml})  reduces to
\begin{multline}\label{e:Mlapprox}
  i\mathcal M^{(l)}_3(p_1,q)\simeq -i^l \left(2\sqrt{2\pi G_N}
    m\right)^{l+1}\cr\times\int \prod_{n=1}^{l}
  {d^d\vec\ell_n\over(2\pi)^d}\prod_{i=1}^{l+1}{\mathcal
    P^{00,\rho_i\sigma_i} \over \prod_{i=1}^{l+1}(\ell_i^2+i\epsilon)}
  \mathcal
    M_{\rho_1\sigma_1,\dots,\rho_{l+1}\sigma_{l+1}} (\ell_1,\dots,\ell_{l+1},q)\Big|_{\ell_i^0=0}
\end{multline}
with $\ell_1+\cdots+\ell_{l+1}=q$.
We recall that
\begin{equation}
  \mathcal P^{00,\rho\sigma}=
  \delta^\rho_0\delta^\sigma_0-{\eta^{\rho\sigma}\over D-2}  \,.
\end{equation}
The amplitude~(\ref{e:Mlapprox})  corresponds to the graph where the scalar line has been collapsed to a point
\begin{equation}
   \mathcal M^{(l)}_3(p_1,q)\simeq 
\begin{gathered}
\begin{fmffile}{gravtreencol}
    \begin{fmfgraph*}(180,100)
%\fmfstraight
\fmfsurroundn{i}{8}
\fmf{phantom,tension=10}{i6,v2,v3,v4,i4}
%\fmf{phantom,tension=10}{i7,v2,v3,v4,v5,i3}
\fmf{dbl_wiggly}{i5,v2}
\fmf{dbl_wiggly}{i5,v3}
\fmf{dbl_wiggly}{i5,v4}
% \fmf{dbl_wiggly,tension=2}{w2,v2}
%  \fmf{dbl_wiggly,tension=2}{w3,v3}
%  \fmf{dbl_wiggly,tension=2}{w4,v4}
%  \fmf{dbl_wiggly,tension=2}{w5,v5}
 \fmfrpolyn{smooth,label={tree},filled=30,tension=.8}{G}{4}
% %\fmffreeze
 \fmf{dbl_wiggly}{G1,v2}
 \fmf{dbl_wiggly}{G2,v3}
 \fmf{dbl_wiggly}{G3,v4}
 \fmf{dbl_wiggly,tension=3}{G4,i1}
\end{fmfgraph*}
\end{fmffile}
\end{gathered}\,.
\end{equation}
In the static with $q=(0,\vec q)$, $|q|\ll m$, the $l+2$-tree level
gravitons amplitude has the leading behaviour 
\begin{equation}
\prod_{n=1}^{l+1}\mathcal P^{00,\rho_i\sigma_i} \mathcal
M_{\rho_i\sigma_i,\cdots,\rho_{l+1}\sigma_{l+1}}(\ell_1,\dots,\ell_{l+1},q)\propto
{\sqrt{G_N}}^l q^2\,,
\end{equation}
and higher powers of $\vec q^2$ contribute to higher powers of $\hslash$
and are sub-leading
quantum corrections (see section~\ref{sec:tree} for more about this).

Therefore, the classical contribution to the stress-tensor
in~\eqref{e:MtoT}  is given by\footnote{We have checked this explicitly to three-loop order
  using the {\tt LiteRed} code~\cite{Lee:2012cn,Lee:2013mka}.}
 \begin{equation}\label{e:Tstatic}
\langle T^{(l)}_{\mu\nu}\rangle=\pi^l(G_Nm)^{l} m\Big(c^{(l)}_1(d)\delta_{\mu}^0\delta_{\nu}^0
+c^{(l)}_2(d)\big({q_{\mu}q_{\nu}\over q^2} -\eta_{\mu\nu}\big) \Big)\,  J_{(l)}(q^2)\,,
\end{equation}
where $c^{(l)}_1(d)$ and $c^{(l)}_2(d)$ are rational functions of the dimension
$d$ and  $J_{(n)}(q^2)$ is the massless $n$-loop sunset graph
\begin{equation}\label{e:Jmastersunset}
  J_{(n)}(\vec q^2)=
  \begin{gathered}\tikzpicture[scale=1.7]
\scope[xshift=-5cm,yshift=-0.4cm,decoration={
    markings,
    mark=at position 0.5 with {\arrow{>}}}]
\draw(0.0,0) node{};
\draw (1,0.0) node{$\bullet$}   ;
\draw (2,0) node{$\bullet$} ;
\draw (0.5,0) node[left]{$q$};
\draw[postaction={decorate}] (0.5,0) -- (1,0);
\draw[dashed] (1,0) -- (2,0) ;
\draw[postaction={decorate}] (2,0) -- (2.5,0);
\draw (2.5,0) node[right]{$q$};
\draw[dashed] (1.5,0.0) ellipse (0.5 and 0.1);
\draw[dashed] (1.5,0.0) ellipse (0.5 and 0.2);
\draw[dashed] (1.5,0.0) ellipse (0.5 and 0.3);
\draw[dashed] (1.5,0.0) ellipse (0.5 and 0.4);
\endscope
\endtikzpicture
\end{gathered}
=\int {\vec
    q^2\over \prod_{i=1}^n \vec l_i^2 \, (\vec l_1+\cdots+\vec
             l_n+\vec q)^2}\prod_{i=1}^n{ d^d{\vec l}_i\over (2\pi)^{d}}  
\,.
\end{equation}

\subsection{The master integrals for the classical limit}
\label{sec:mast-integr-class}

The master integrals~(\ref{e:Jmastersunset})   can be evaluated
straightforwardly  with the
 parametric representation of the $n$-loop sunset  in $D$ dimensions
 (see~\cite{Vanhove:2014wqa})
 \begin{equation}
J_{(n)}(\vec q^2)   ={ (\vec q^2)^{n(d-2)\over2} \over
  (4\pi)^{nd\over2}}\Gamma\left(n+1 -{n d\over 2} \right)
\int_{x_i\geq0}  \left({1\over x_1}+\cdots +{1\over x_n}+1\right)^{(n+1)(2-d)\over2}
\prod_{i=1}^{n} {dx_i   \over x_i^{d\over 2}}
 \end{equation}
 since the first Symanzik polynomial is $U_{n+1}=
\left( \sum_{i=1}^{n+1}{1\over x_i}\right)\left(\prod_{i=1}^{n+1} x_i\right)$
 and the second Symanzik polynomial is $F_{n+1}= -q^2 x_1\cdots
 x_{n+1}=\vec q^2 x_1\cdots x_{n+1}$.
 Changing variables to  $y_i=1/x_i$  we have
 \begin{equation}
J_{(n)}(\vec q^2)   ={ (\vec q^2)^{n(d-2)\over2} \over
  (4\pi)^{nd\over2}}\Gamma\left(n+1 -{n d\over 2} \right)
\int_{y_i\geq0}  \left(y_1+\cdots + y_n+1\right)^{(n+1)(2-d)\over2}
\prod_{i=1}^{n} {dy_i   \over y_i^{4-d\over 2}}\,.
 \end{equation}
 Using the expression for Euler's beta-function
 \begin{equation}
   \int_0^\infty  (x+a)^{\alpha}
{dx   \over x^{1-\beta}}=  a^{\alpha+\beta}
{\Gamma(-\beta-\alpha)\Gamma(\beta)\over \Gamma(-\alpha)},
 \end{equation}
 the master integral is readily evaluated to be
 \begin{equation}\label{e:Jnresult}
J_{(n)}(\vec q^2)  ={ (\vec q^2)^{n(d-2)\over2} \over
  (4\pi)^{nd\over2}}
{\Gamma\left(n+1 -{n d\over 2} \right)  \Gamma\left(d-2\over2\right)^{n+1}\over \Gamma\left((n+1)(d-2)\over2\right)}\,.
\end{equation}
The master integrals develop  ultraviolet poles at loop orders,
inducing divergences in the stress-energy tensor. We
will show in section~\ref{sec:nonmin} how to renormalise these divergences with the
introduction of higher-derivative couplings.

\section{The metric perturbation from graviton emission}
\label{sec:class-metr-pert}

Using the relation~(\ref{e:TtohAmplitudedeDonder}) between the metric
perturbation and using the expression~(\ref{e:Tstatic}) for  the stress-energy tensor in $d$-dimension  in the
static limit we have
\begin{multline}
  h_{\mu\nu}^{(l+1)}(\vec q)= -8
  \left(c_1^{(l)}(d)(2\delta^0_\mu\delta^0_\nu-\eta_{\mu\nu})+c_2^{(l)}(d)\left(2{q_\mu
      q_\nu\over q^2}+(d-2)\eta_{\mu\nu}\right)\right)\cr
\times { (\pi  G_Nm)^{l+1}J_{(l)}(\vec q^2) \over
    \vec q^2}\,.
\end{multline}
The static space-time components are obtained by computing the Fourier
transform in $d$  dimensions
\begin{equation}
 h^{(l+1)}_{\mu\nu}(\vec x) = \int_{\mathbb R^d}  h_{\mu\nu}^{(l+1)}(\vec q)e^{i\vec q\cdot
   \vec x} {d^d{\vec q}\over(2\pi)^d} \,.
\end{equation}
Using the Fourier transformations given in appendix~\ref{sec:FT}, and 
setting $r=|\vec
x|$, the Fourier transform of the master integrals are given by
\begin{equation}\label{e:Jnr}
\int_{\mathbb R^d}  {J_{(l)}(\vec q^2)\over \vec q^2} e^{i\vec
    q\cdot \vec x} {d^d\vec q\over (2\pi)^d}
  =\left({\Gamma\left(d-2\over2\right)\over
    4\pi^{d\over2}} {1\over r^{d-2}}\right)^{l+1}
\end{equation}
which is finite to all loop orders. The
ultraviolet divergences in the momentum space representation
in~\eqref{e:Jnresult} has been cancelled by the Fourier transform.\footnote{This fact had been noticed by L. Plant\'e
  in his PhD thesis~\cite{PlanteThesis}.}

The tensorial Fourier transform
\begin{equation}\label{e:Jnrxx}
  \int_{\mathbb R^d} {q_iq_j\over \vec q^2} {J_{(l)}(\vec q^2)\over \vec q^2} e^{i\vec
    q\cdot \vec x} {d^d\vec q\over (2\pi)^d}
  =\left({\Gamma\left(d-2\over2\right)\over
    4\pi^{d\over2}} {1\over r^{d-2}}\right)^{l+1}{1\over2-l(d-2)}\left(
   - \delta_{ij}+ (l+1)(d-2) {x_ix_j\over r}\right)\,.
\end{equation}
diverges for $l=1$ and $d=4$ and for  $l=2$ and $d=3$, and are
otherwise finite.

By spherical symmetry we parameterise the metric  in $d+1$ dimensions
\begin{equation}
ds^2=h_0(r,d) dt^2-  h_1(r,d) d\vec x^2-h_2(r,d) {(\vec           x\cdot
  d\vec x)^2\over \vec x^2}\,,
\end{equation}
so that
\begin{equation}
  h_{i}(\vec x)=h_i^{(0)}+\sum_{l\geq1} h_i^{(l)}(\vec x)\,,
\end{equation}
with $h_i^{(0)}=1,1,0$ for $i=0,1,2$, 
the post-Minkowskian expansion of the metric components
\begin{align}\label{e:h0amp}
  h^{(l+1)}_{0}(r,d)&=-\frac{16}{d-1} \left((d-2)c^{(l)}_1(d)+ c^{(l)}_2(d) \right)
  \left(\rho(r,d)\over 4\right)^{l+1} ,\\
  h^{(l+1)}_{1}(r,d)&=\frac{16}{d-1}
 \left(c^{(l)}_1(d)-\left(1+{d-1\over 2-l(d-2)}\right)c^{(l)}_2(d)\right)  \left(\rho(r,d)\over 4\right)^{l+1}  ,
 \cr
\nonumber h^{(l+1)}_2(r,d)&= 16 {(d-2)(l+1)\over2-l(d-2)}  c^{(l)}_2(d)
  \left(\rho(r,d)\over 4\right)^{l+1}    \,.
\end{align}
We have introduced the radial parameter
\begin{equation}\label{e:rhodef}
  \rho(r,d)={\Gamma\left(d-2\over2\right)\over \pi^{d-2\over 2} } {
      G_N m \over r^{d-2}}\,,
\end{equation}
which is our post-Minkowskian expansion parameter.   Recall that in
$d+1$ dimensions the length dimension of $[G_N m]= (length)^{d-2}$
and $\rho(r,d)$ is dimensionless.

The metric component  present poles in four dimensions ($d=3$) from
two-loop order and in five
dimensions ($d=4$) from one-loop order. Such divergences will be
removed by the contribution from the non-minimal coupling contributions in section~\ref{sec:nonmin}.

%------------------------------------------------------------------------
\subsection{Tree-level amplitude}\label{sec:tree}
At tree-level, the only contributing diagram  is

\begin{equation}
\mathcal M^{(0)}_3(p_1,q)  =
\begin{gathered}\begin{fmffile}{gravtree}
    \begin{fmfgraph*}(80,100)
\fmfstraight
\fmfleftn{i}{2}
\fmfrightn{o}{1}
\fmf{fermion,tension=2.5,label=$p_1$}{i1,v1}
\fmf{fermion,tension=2.5,label=$p_2$}{v1,i2}
\fmf{dbl_wiggly,tension=1, label.dist=10,label=$q$}{o1,v1}
\end{fmfgraph*}
\end{fmffile}
\end{gathered}\,,
\end{equation}
 is the emission of a graviton
from the scattering of two massive scalars of momenta $p_1$ and $p_2$
and $p_1^2=p_2^2=m^2$  with momentum transfert
$q=p_1-p_2$. The scattering amplitude is given by the
2-scalar-1-graviton vertex $\tau^{\mu\nu}(p_1,p_2)$ in~(\ref{e:tau1})
\begin{equation}
i \mathcal M ^{(0)}_3(p_1,q)=-{i\sqrt{32\pi G_N}\over2\sqrt{4E_1E_2}} \epsilon^{\mu\nu}
 \tau_{\mu\nu}  =- {i\sqrt{32\pi G_N}\over2} \epsilon^{\mu\nu}
 \big(p_{1\mu}p_{2\nu}+p_{2\mu}p_{1\nu}-\eta_{\mu\nu}(p_1\cdot p_2-m^2)\big)\,.
\end{equation}
Using that $P=(p_1+p_2)/2$ and $q=p_1-p_2$ we have that
\begin{equation}
 i\mathcal M ^{(0)}_3(p_1,q)=- {i\sqrt{32\pi G_N}\over2\sqrt{4E_1E_2}} \epsilon^{\mu\nu}
 \big(2 P_{\mu}P_{\nu}-\frac12 (q_\mu q_\nu-\eta_{\mu\nu}q^2)\big)\,.
\end{equation}
In the static limit $q=p_1-p_2\simeq(0,\vec q)$, $E_1\simeq E_2\simeq m$  and $|\vec q|\ll m$
we have 
\begin{equation}
  \label{eq:1}
 \langle T^{(0)}_{\mu\nu}(q^2)\rangle \simeq    m
\delta^0_\mu\delta^0_\nu +\left({q_iq_j\over \vec
    2q^2}\eta^i_\mu\eta^j_\nu+\frac12\eta_{\mu\nu}\right) \vec q^2\,.
\end{equation}
The $\vec q^2$ term in this expression is the contact term which has
a higher power of $\hslash$ and does not contribute to the classical
limit~\cite{Bjerrum-Bohr:2018xdl,Cristofoli:2019neg}.
The coefficients of the classical contribution to the stress-tensor at
tree-level are given by 
\begin{align}\label{e:F1F2treeAmp}
c_1^{(0)}(d)&= 1,\cr
c_2^{(0)}(d)&= 0\,.
\end{align}
From this we deduce the metric components in $d+1$ dimensions
using~(\ref{e:h0amp}) 
\begin{align}\label{e:htree}
  h_{0}^{(1)}(r,d)&= -4{d-2\over d-1} \rho(r,d),\cr
h_{1}^{(1)}(r,d)&={4\over d-1}\rho(r,d), \cr
h_{2}^{(1)}(r,d)&=0\,,                
\end{align}
where   $\rho(r,d)$ is defined   in~(\ref{e:rhodef}). This reproduces
the expression given in~\cite{Collado:2018isu,Jakobsen:2020ksu}. 
%------------------------------------------------------------------------
\subsection{One-loop amplitude}\label{sec:oneloop}

At one-loop  the only contributing diagram to the classical limit 
is
\begin{equation}
i\mathcal M^{(1)}_3(p_1,q)  =\begin{gathered}\begin{fmffile}{oneloopgraph}
  \begin{fmfgraph*}(100,100)
      \fmfleftn{i}{2}
      \fmfrightn{o}{1}
\fmf{fermion,label=$p_1$}{i1,v1}
\fmf{fermion,label=$p_2$}{v2,i2}
\fmf{dbl_wiggly,label=$q$}{o1,v3}
\fmf{plain,tension=.1}{v1,v2}
\fmf{dbl_wiggly,tension=.3}{v3,v1}
\fmf{dbl_wiggly,tension=.3}{v3,v2}
\end{fmfgraph*}
\end{fmffile}
\end{gathered}=-{i\sqrt{32\pi G_N}\over2} \epsilon_{\mu\nu}
 T^{(1)\ \mu\nu}(q^2)\,,
\end{equation}
from which we extract the one-loop contribution to the stress-energy
tensor in $d+1$ dimensions
\begin{equation}
 T^{(1)\ \mu\nu}(q^2)=\frac{i 8\pi G_N}{\sqrt{4
     E_1E_2}}\int\frac{d^{d+1}l}{(2\pi)^D}\frac{\tau^{\sigma\rho}(p_1,l+p_1)  \tau^{\mu\nu}_{(3)\sigma\rho,\kappa\delta}(l,q)
   \tau^{\kappa\delta} (p_2,l+p_1)
  }{(l^2+i\epsilon)
   ((l+q)^2+i\epsilon) ((l+p_1)^2-m^2+i\epsilon)},
\end{equation}
where $\tau^{\mu\nu}_{(3)\
  \pi\rho,\sigma\tau}(p_1,p_2)$ is the three graviton vertex  and $\tau^{\mu\nu}(p_1,p_2)$ the
vertex for the emission of a graviton from two scalars with momenta
$p_1$ and $p_2$. We refer to appendix~\ref{sec:vertices} for
definitions and normalisation of our vertices.

 In the static limit, $\vec{q}^2\ll m^2$, the classical contribution
 coming from the two scalars to one-graviton vertex is 
\begin{equation}\label{e:tau1limit}
\tau_{\alpha\beta}\approx 2 m^2\delta^0_{\alpha}\delta^0_{\beta},
\end{equation}
using that $p_1^2=p_2^2=m^2$.
This gives for the stress-energy tensor
\begin{equation}
T^{(1)\
  \mu\nu}(q^2)=i 16 \pi G_N m^3\int\frac{d^{d+1}l}{(2\pi)^D}\frac{ \tau^{\mu\nu}_{(3)00,00}(l,q)}{(l^2+i\epsilon)
  ((l+q)^2+i\epsilon) ((l+p_1)^2-m^2+i\epsilon)}.
\end{equation}
At this point, we want to focus on the computation of the classical contribution at the static limit. Thus, we will employ a trick, which will prove useful for higher loops. We symmetrize the diagram
\begin{multline}
T^{(1)\
  \mu\nu}(q^2)=i 8 \pi G_N m^3\int\frac{d^{d+1}l}{(2\pi)^D}\frac{ \tau^{\mu\nu}_{(3)00,00}(l,q)}{(l^2+i\epsilon)
  ((l+q)^2+i\epsilon)}\cr
\times\Big[\frac{1}{(l+p_1)^2-m^2+i\epsilon}+\frac{1}{(l-p_2)^2-m^2+i\epsilon}\Big]\,.
\end{multline}

In the approximation  $l^2\ll m^2$ we have  $(l+p_i)^2-m^2=l^2+2l\cdot
p_1=l^2+2l_{0}E-\vec{l}\cdot\vec{q}\simeq l_0^2 +2ml_0$ and the
amplitude reduces at leading order
 \begin{multline}
T^{(1)\ \mu\nu}(q^2)\simeq i8\pi G_N
m^3\int\frac{d^{d+1}l}{(2\pi)^D}\frac{ \tau^{\mu\nu}_{(3)00,00}(l,q)}{(l^2+i\epsilon)((l+q)^2+i\epsilon)}\cr
\times\Big[\frac{1}{l_0^2+2ml_0+i\epsilon}+\frac{1}{l_0^2-2ml_0+i\epsilon}\Big].
\end{multline}
It is obvious that at $\mathcal{O}(\epsilon^0)$ order we get a zero
contribution at leading order in $1/m$, since $l_0\ll m$. Thus, we can
compute the leading contribution of the integral over $l_0$ via
\textit{Cauchy's theorem}, by taking the residue
$2ml_0=i\epsilon$ and closing the contour of integration in the upper
half-plane\footnote{One could have taken the residue at
  $2ml_0=-i\epsilon$ and closing the contour in the lower half-plane
  with the same result.}
 
 \begin{equation}
T^{(1)\ \mu\nu}(q^2)=4\pi G_N m^2\int\frac{d^d{\vec l}}{(2\pi)^{d}}\frac{ \tau^{\mu\nu}_{(3)00,00}(l,q)}{(\vec{l}^2-i\epsilon)((\vec{l}+\vec{q})^2-i\epsilon)}\Bigg\vert_{l_0=0}\,,
\end{equation}
with
\begin{align}
 \tau^{\mu\nu}_{(3)00,00}(l,q)=\frac{1}{d-1}\bigg(&(d-2)\big(l^{\mu}l^{\nu}+(l+q)^{\mu}(l+q)^{\nu}+q^{\mu}q^{\nu}+\frac{3}{2}\eta^{\mu\nu}\vec{q}^2\big)  \nn \\
 &-2(d-2)\big(\vec{l_1}^2+(\vec{l_1}+\vec{q})^2\big)(\delta^{\mu}_0\delta^{\mu}_0-\frac{\eta^{\mu\nu}}{4})-2(d-3) \vec{q}^2\delta^{\mu}_0\delta^{\mu}_0\bigg)\,.
\end{align}
The component of the stress-tensor are proportional to the one-loop master
integral $J_{(1)}(\vec q^2)$ as expected from the general discussion
of section~\ref{sec:mast-integr-class}

\begin{equation}
  \label{e:Toneloop}
  \langle  T_{\mu\nu}^{(1)}\rangle= \pi G_Nm^2
  \left(c_1^{(1)}(d) \delta_\mu^0\delta_\nu^0+ c_2^{(1)}(d)
    \left({q_\mu q_\nu\over q^2}-\eta_{\mu\nu}\right)\right)\, J_{(1)}(q^2)\,, 
\end{equation}
with the master integral
\begin{equation}
  \label{e:J1}
  J_{(1)}(q^2)=\frac{ \Gamma \left(4-d\over2\right) \Gamma
   \left(\frac{d-2}{2}\right)^2}{2^d\pi^{d\over2}\Gamma (d-2)}\,
 \left(\vec q^2\right)^{\frac{d-2}{2}}\,,
\end{equation}
and the coefficients 
 \begin{align}
  c_1^{(1)}(d)&=-\frac{2(4d^2-15d+10)}{(d-1)^2},\cr
  c_2^{(1)}(d)&=-\frac{2(d-2)(3d-2)}{(d-1)^2}\,.
 \end{align}

%---------------------------------------------------------------------
 \subsubsection{The one-loop contribution to the metric components}
\label{sec:one-loop-metric}
Using~\eqref{e:h0amp} we get for the metric components in $d+1$ dimensions
\begin{align}\label{e:honeloop}
  h_{0}^{(2)}(r,d)&={8(d-2)^2\over (d-1)^2}\rho(r,d)^2,\cr
h_1^{(2)}(r,d)&=-\frac{4(2d^2-9d+14)}{(d-4) (d-1)^2}\rho(r,d)^2,\cr
 h_2^{(2)}(r,d)&=\frac{4 (d-2)^2(3d-2)}{(d-4) (d-1)^2}\rho(r,d)^2 \,,
\end{align}
where  $\rho(r,d)$ is defined in~(\ref{e:rhodef}).

This reproduces the expression given in~\cite{Collado:2018isu} and the
expression in~\cite[eq.~(22)]{Jakobsen:2020ksu} for $\alpha=0$.

%---------------------------------------------------------------------
\subsection{Two-loop amplitude}\label{sec:twoloop}

The diagrams contributing to the classical corrections at third post-Minkowskian
order of the metric at the two-loop graphs 
\begin{equation}
i  \mathcal M^{(2)}_3(p_1,q)= -\sqrt{32 \pi G_N}  T^{(2)\, \mu\nu}  \epsilon_{\mu\nu},
\end{equation}
there are four contributions
\begin{align}
T_{(a)}^{(2)\mu\nu}&=
\begin{gathered}
\begin{fmffile}{twolooptriangle4pt12}
 \begin{fmfgraph}(100,50)
\fmfstraight
      \fmfleftn{i}{2}
      \fmfrightn{o}{1}
\fmf{plain,tension=10}{i1,v1,v4,v2,i2}
\fmffreeze
\fmf{dbl_wiggly,tension=3}{v3,v5,v2}
\fmf{dbl_wiggly}{v3,v1}
\fmf{dbl_wiggly,tension=.1}{v5,v4}
\fmf{dbl_wiggly,tension=5}{v3,o1}
\end{fmfgraph}
\end{fmffile}
\end{gathered},\qquad 
\nonumber T_{(b)}^{(2)\mu\nu}=
\begin{gathered}
\begin{fmffile}{twolooptriangle4pt21}
  \begin{fmfgraph}(100,50)
\fmfstraight
      \fmfleftn{i}{2}
      \fmfrightn{o}{1}
\fmf{plain,tension=10}{i1,v1,v4,v2,i2}
\fmffreeze
\fmf{dbl_wiggly,tension=3}{v3,v5,v1}
\fmf{dbl_wiggly}{v3,v2}
\fmf{dbl_wiggly,tension=.1}{v5,v4}
\fmf{dbl_wiggly,tension=5}{o1,v3}
\end{fmfgraph}
\end{fmffile}
\end{gathered},\\
\nonumber 
T_{(c)}^{(2)\mu\nu}&=
\begin{gathered}
\begin{fmffile}{twolooptriangle3}
  \begin{fmfgraph}(100,50)
\fmfstraight
      \fmfleftn{i}{2}
      \fmfrightn{o}{1}
\fmf{plain}{i1,v1,vph1a,vph1b,vph1c,v4,vph2,v2,i2}
\fmffreeze
\fmf{dbl_wiggly}{v3,v1}
\fmf{dbl_wiggly}{v3,v2}
\fmf{dbl_wiggly,tension=3}{o1,v5,v3}
\fmf{dbl_wiggly,left=.5,tension=.1}{v5,v4}
\end{fmfgraph}
\end{fmffile}
\end{gathered},\qquad
T_{(d)}^{(2)\mu\nu}=\begin{gathered}
 \begin{fmffile}{twolooptriangle4pt}
  \begin{fmfgraph}(100,50)
\fmfstraight
      \fmfleftn{i}{2}
      \fmfrightn{o}{1}
\fmf{plain,tension=10}{i1,ov1,ovph1,ovph2,ov3,ovph3,ovph4,ov2,i2}
\fmffreeze
\fmf{dbl_wiggly,tension=.3}{v3,ov1}
\fmf{dbl_wiggly,tension=.3}{v3,ov2}
\fmf{dbl_wiggly,tension=.3}{v3,ov3}
\fmf{dbl_wiggly,tension=1}{o1,v3}
\end{fmfgraph}
\end{fmffile}
\end{gathered}\,.
\end{align}

\subsubsection{The diagrams $(a)$, $(b)$, $(c)$}
\label{sec:diagrams-abc}
The sum of the contributions from the diagrams $(a)$, $(b)$, $(c)$
 after appropriate labelling of the momenta, can be expressed as
\begin{multline}
\sum_{i=a}^c T_{(i)}^{(2)\,\mu\nu}=-{16 G_N^2\pi^2\over m}\int
\prod_{n=1}^3\frac{d^{d+1}l_n}{(2\pi)^{2d}}\delta(l_1+l_2+l_3+q)\cr
\times\frac{\tau^{\gamma\delta} (p_1,l_1+p_1) 
\tau ^{\sigma\tau} (l_1+p_1,-l_2+p_1) 
\tau ^{\iota\theta} (l_2-p_2,-p_2)
              \tau^{\phi\chi}_{(3)\iota\theta,\sigma\tau}(-l_2,l_1+q)\cdot\mathcal{P}^{\alpha\beta}_{\phi\chi}\cdot
               \tau_{(3)\alpha\beta,\gamma\delta}^{\mu\nu}(l_1+q,q)}{l_1^2l_2^2l_3^2(l_1+q)^2}\cr
\times\Bigg(\frac{1}{(l_1+p_1)^2-m^2}\frac{1}{(l_2-p_2)^2-m^2}+\frac{1}{(l_3+p_1)^2-m^2}\frac{1}{(l_1-p_2)^2-m^2}\cr+\frac{1}{(l_3+p_1)^2-m^2}\frac{1}{(l_2-p_2)^2-m^2}  \Bigg).      
\end{multline}
Using the approximate form of the two scalars one graviton vertex
in~(\ref{e:tau1limit}) and $(l_1+p_1)^2-m^2\approx 2ml_1^0$ and taking the residue $2ml_i^0= i\epsilon$, since for the rest of the residues we get a zero contribution at order $\mathcal{O}(\epsilon^0)$, we get

\begin{equation}
\sum_{i=a}^c T_{(i)}^{(2)\,\mu\nu}=32\pi^2G_N^2 m^3\int \prod_{n=1}^2\frac{d^{d+1}l_n}{(2\pi)^{2d}}\frac{\tau_{(3) \alpha\beta,00}^{\mu\nu}(l_1+q,q)\cdot\mathcal{P}^{\alpha\beta}_{\phi\chi}\cdot\tau_{(3)\, 00,00}^{\phi\chi}(-l_2,l_1+q)}{(\vec{l_1})^{^2}(\vec{l_2})^{^2}(\vec{l_3})^{^2}(\vec{l_1}+\vec{q})^{^2}}\Bigg\vert_{l_1^0=l_2^0=0},
\end{equation}
with
\begin{multline}
 \eta_{\mu\nu} \tau_{(3) \phi\chi,00}^{\mu\nu}(l_1+q,q)=\big(l^{\mu}l^{\nu}-(l+q)^{\mu}(l+q)^{\nu}-q^{\mu}q^{\nu}\big)-\frac{3}{2}\eta^{\mu\nu}\vec{q}^2\big(\eta^{\mu\nu}-(d-1)\delta^{\mu}_0\delta^{\nu}_0\big)  \cr
 +\frac{\eta^{\mu\nu}}{2}\big(\vec{l_1}^2-(\vec{l_1}+\vec{q})^2\big)-\frac{5-d}{2}\delta^{\mu}_0\delta^{\nu}_0\big(\vec{l_1}^2+(\vec{l_1}+\vec{q})^2\big),
\end{multline}
and
\begin{multline}
 \delta_{\mu}^0\delta_{\nu}^0\tau_{(3) \phi\chi,00}^{\mu\nu}(l_1+q,q)=\frac{1}{d-1}\bigg((d-3)\big((l+q)^{\mu}(l+q)^{\nu}+q^{\mu}q^{\nu}\big)+(d-1)\big(l_1^{\mu}l_1^{\nu}-\frac{\vec{l_1}^2}{2}(3\delta^{\mu}_0\delta^{\nu}_0-\eta^{\mu\nu}) \big)  \cr
 +\frac{\eta^{\mu\nu}-\delta^{\mu}_0\delta^{\nu}_0}{2}\big(\vec{q}^2(d-5)+(3d-7)(\vec{l_1}+\vec{q})^2\big)\bigg),
\end{multline}
  and
\begin{multline}
 \tau^{\mu\nu}_{(3)00,00}(l,q)=\frac{1}{d-1}\bigg((d-2)\big(l^{\mu}l^{\nu}+(l+q)^{\mu}(l+q)^{\nu}+q^{\mu}q^{\nu}+\frac{3}{2}\eta^{\mu\nu}\vec{q}^2\big)  \cr
 -2(d-2)\big(\vec{l_1}^2+(\vec{l_1}+\vec{q})^2\big)(\delta^{\mu}_0\delta^{\mu}_0-\frac{\eta^{\mu\nu}}{4})-2(d-3) \vec{q}^2\delta^{\mu}_0\delta^{\mu}_0\bigg).
\end{multline}
Using the {\tt LiteRed} code~\cite{Lee:2012cn,Lee:2013mka} in $d$ dimensions, we
find that all the contributions are proportional to the master
integral as expected from the general discussion of
section~\ref{sec:mast-integr-class}
\begin{align}
  J_{(2)}(\vec q)&= \int \prod_{i=1}^2 {d^d{\vec l}_i\over(2\pi)^d}  {\vec
    q^2\over \prod_{i=1}^2 \vec l_i^2 (\vec l_1+\vec l_2+\vec q)^2}\cr
  &=
  -{\vec q^2\over 32\pi^2(d-3)}-\left(-3+\gamma_E-\log(4\pi)+\log(\vec
    q^2)\right)\vec q^2+O(d-3)\,,
\end{align}
where $\gamma_E=0.57721\cdots$ is the Euler-Mascheroni constant~\cite{Lagarias}.

We find for the 00-component 
\begin{equation}
\sum_{i=a}^cT_{(i)}^{(2)\,00}=\frac{32\pi^2G_N^2m^3}{3}\frac{6d^3-45d^2+134d-160}{(d-4)(d-1)^2} J_{(2)}(\vec q^2)\,,
\end{equation}
and for the trace part
\begin{equation}
\sum_{i=a}^cT_{(i)}^{(2)\,\mu\nu}\eta_{\mu\nu}=-\frac{32\pi^2
  G_N^2m^3}{3}\frac{10d^3-63d^2+123d-86}{(d-1)^2}J_{(2)}(\vec q^2)\,.
\end{equation}

\subsubsection{The diagrams $(d)$}
\label{sec:diagrams-d}
The diagram $(d)$ after symmetrisation over the massive scalar legs reads
\begin{multline}
T_{(d)}^{(2)\,\mu\nu}=-\frac{ 32 G_N^2\pi^2}{3m}\int \prod_{n=1}^3\frac{d^{d+1}l_n}{(2\pi)^{2d}}\frac{\delta(l_1+l_2+l_3+q)}{l_1^2l_2^2l_3^2}
\Bigg(\frac{1}{(l_1+p_1)^2-m^2+i\epsilon}\frac{1}{(l_2-p_2)^2-m^2+i\epsilon}\cr+\frac{1}{(l_3+p_1)^2-m^2+i\epsilon}\frac{1}{(l_1-p_2)^2-m^2+i\epsilon}+\frac{1}{(l_3+p_1)^2-m^2+i\epsilon}\frac{1}{(l_2-p_2)^2-m^2+i\epsilon}
 \Bigg)\cr
\times  \tau ^{\gamma\delta} (p_1,l_1+p_1)\tau ^{\sigma\tau} (l_1+p_1,-l_2+p_1)\tau  ^{\iota\theta} (l_2-p_2,-p_2)\tau_{(4)\gamma\delta,\sigma\tau,\iota\theta}^{\mu\nu}(q,l_1,l_2,l_3),
\end{multline}
and leads to the contribution

\begin{equation}
T_{(d)}^{(2)\,\mu\nu}=-\frac{64\pi^2G_N^2 m^3}{3}\int
\prod_{n=1}^2\frac{d^{d+1}l_n}{(2\pi)^{d}}\frac{\tau_{(4)\,00,00,00}^{\mu\nu}(q,l_1,l_2,-l_1-l_2-q)}{(\vec{l_1})^{^2}(\vec{l_2})^{^2}(\vec{l_1}+\vec{l_2}+\vec
 {q})^{^2}}\Bigg\vert_{l_1^0=l_2^0=0}
\end{equation}
with the vertex 
\begin{multline}
    \tau_{(4)\,00,00,00}^{\mu\nu}(q,l_1,l_2,l_3)=\frac{1}{(d-1)^2}\Bigg(\vec{q}^2\frac{\delta^{\mu}_0\delta^{\nu}_0}{2}
    (7d^2-45d+70)-\vec{q}^2\frac{\eta^{\mu\nu}}{2}(d-2)(6d-23) \cr
    +(d-2)\bigg((9-2d)q^{\mu}q^{\nu}+(7-2d)\big(l_1^{\mu}l_1^{\nu}+l_2^{\mu}l_2^{\nu}+l_3^{\mu}l_3^{\nu}\big)\bigg) \cr
    +\frac{d-2}{2}\big(\vec{l_1}^2+\vec{l_2}^2+\vec{l_3}^2\big) \big(\delta^{\mu}_0\delta^{\nu}_0(7d-23)-\eta^{\mu\nu}(2d-9)\big)   \Bigg)\,.
\end{multline}
Evaluating these integral we find, for  the $00$-component

\begin{equation}
T_{(d)}^{(2)\,00}=-\frac{32\pi^2G_N^2m^3}{3} \frac{(4-d)(6-d)}{(d-1)^2}J_{(2)}(\vec q^2)\,,
\end{equation}
and for the trace part
\begin{equation}
T_{(d)}^{(2)\,\mu\nu}\eta_{\mu\nu}=\frac{64\pi^2G_N^2m^3}{3} \frac{3d^3-20d^2+41d-30}{(d-1)^2}J_{(2)}(\vec
q^2)\,.
\end{equation}

\subsubsection{The two-loop contribution to the metric components}
Summing up all the contributions   the two-loop stress-tensor is given by
\begin{equation}\label{e:Tstatictwoloop}
\langle T^{(2)}_{\mu\nu}\rangle=\pi^2 G_N^2 m^3\Big(c^{(2)}_1(d)\delta_{\mu}^0\delta_{\nu}^0
+c^{(2)}_2(d)\big({q_{\mu}q_{\nu}\over q^2} -\eta_{\mu\nu}\big) \Big)\,  J_{(2)}(q^2)\,,
\end{equation}
with the coefficients given by
\begin{align}\label{e:c1c22loop}
c_1^{(2)}(d) &={32\over3(d-4)(d-1)^3}\left(9d^4-70d^3+203d^2-254d+104  \right),\cr
c_2^{(2)}(d) &={64(d-2)\over3(d-4)(d-1)^3}\left(2d^3-13d^2+25d-10 \right)\,,
\end{align}
and the expression for the master integral
\begin{equation}\label{e:J2}
  J_{(2)}(\vec q^2)=\frac{\Gamma (3-d) \Gamma
    \left(\frac{d-2}{2}\right)^3}{(4\pi)^d\Gamma \left(\frac{3
        (d-2)}{2}\right)}\left(\vec q^2\right)^{d-2}\,.  
\end{equation}
From which we extract the metric components using the relations~(\ref{e:h0amp}) (using the
definition of $\rho(r,d)$ in~(\ref{e:rhodef}))
\begin{align}\label{e:htwoloopdiv}
  h_{0}^{(3)}(r,d)& =-{8(3d-7)(d-2)^3\over (d-4)(d-1)^3}\rho(r,d)^3,\cr
h_1^{(3)}(r,d)&=\frac{8 (7d^4-63d^3+214d^2-334d+212)}{3
                (d-3)(d-4)(d-1)^3}\rho(r,d)^3,\cr
 h_{2}^{(3)}(r,d)&=-\frac{8(d-2)^2(2d^3-13d^2+25d-10)}{
                  (d-3)(d-4)(d-1)^3}\rho(r,d)^3\,.
\end{align}

%-------------------------------------------------------------------------
\subsection{Three-loop amplitude}\label{sec:threeloop}

The diagrams contributing to the classical corrections at third post-Minkowskian
order of the metric at the two-loop graphs 
\begin{equation}
 i \mathcal M^{(3)}_3(p_1,q)=- \sqrt{32 \pi G_N}  T^{(3)\, \mu\nu}  \epsilon_{\mu\nu},
\end{equation}
where the three-loop stress-tensor is given by five distinct diagrams

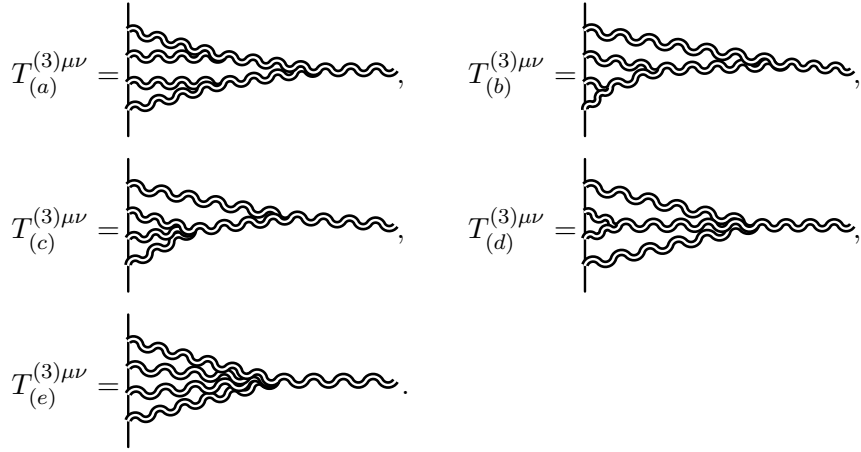
\begin{align}
T_{(a)}^{(3)\mu\nu}&=
\begin{gathered}
\begin{fmffile}{threeloopDa}
 \begin{fmfgraph}(100,50)
\fmfstraight
      \fmfleftn{i}{2}
      \fmfrightn{o}{1}
\fmf{plain,tension=10}{i1,v1,v2,v3,v4,i2}
\fmffreeze
\fmf{dbl_wiggly}{v3,v6,v4}
\fmf{dbl_wiggly}{v2,v5,v1}
\fmf{dbl_wiggly,tension=2}{v5,v7,v6}
\fmf{dbl_wiggly,tension=5}{v7,o1}
\end{fmfgraph}
\end{fmffile}
\end{gathered},\qquad 
\nonumber T_{(b)}^{(3)\mu\nu}=
\begin{gathered}
\begin{fmffile}{threeloopDb}
  \begin{fmfgraph}(100,50)
\fmfstraight
      \fmfleftn{i}{2}
      \fmfrightn{o}{1}
\fmf{plain,tension=10}{i1,v1,v2,v3,v4,i2}
\fmffreeze
\fmf{dbl_wiggly}{v2,v5,v1}
\fmf{dbl_wiggly}{v3,v6,v5}
\fmf{dbl_wiggly,tension=1}{v6,v7,v4}
\fmf{dbl_wiggly,tension=5}{v7,o1}
\end{fmfgraph}
\end{fmffile}
\end{gathered},\\
\nonumber 
T_{(c)}^{(3)\mu\nu}&=
\begin{gathered}
\begin{fmffile}{threeloopDc}
  \begin{fmfgraph}(100,50)
\fmfstraight
      \fmfleftn{i}{2}
      \fmfrightn{o}{1}
\fmf{plain,tension=10}{i1,v1,v2,v3,v4,i2}
\fmffreeze
\fmf{dbl_wiggly}{v5,v1}
\fmf{dbl_wiggly}{v5,v2}
\fmf{dbl_wiggly}{v5,v3}
\fmf{dbl_wiggly,tension=2}{v4,v6,v5}
\fmf{dbl_wiggly,tension=5}{v6,o1}
\end{fmfgraph}
\end{fmffile}
\end{gathered},\qquad
T_{(d)}^{(3)\mu\nu}=\begin{gathered}
 \begin{fmffile}{threeloopDd}
  \begin{fmfgraph}(100,50)
\fmfstraight
      \fmfleftn{i}{2}
      \fmfrightn{o}{1}
\fmf{plain,tension=10}{i1,v1,v2,v3,v4,i2}
\fmffreeze
\fmf{dbl_wiggly}{v6,v1}
\fmf{dbl_wiggly}{v6,v4}
\fmf{dbl_wiggly}{v6,v5}
\fmf{dbl_wiggly,tension=2}{v3,v5,v2}
\fmf{dbl_wiggly,tension=5}{v6,o1}
\end{fmfgraph}
\end{fmffile}
\end{gathered},\\
\nonumber 
T_{(e)}^{(3)\mu\nu}&=
\begin{gathered}
\begin{fmffile}{threeloopDe}
  \begin{fmfgraph}(100,50)
\fmfstraight
      \fmfleftn{i}{2}
      \fmfrightn{o}{1}
\fmf{plain,tension=10}{i1,v1,v2,v3,v4,i2}
\fmffreeze
\fmf{dbl_wiggly}{v5,v1}
\fmf{dbl_wiggly}{v5,v2}
\fmf{dbl_wiggly}{v5,v3}
\fmf{dbl_wiggly}{v5,v4}
\fmf{dbl_wiggly,tension=5}{v5,o1}
\end{fmfgraph}
\end{fmffile}
\end{gathered}\,.
  \end{align}

As before, we permute the internal momenta such that by taking
the residue at $2ml_i^0=i\epsilon$ from the massive propagators, we
extract the non-analytic terms which contribute to the classical
metric in the static limit.
After taking the residues and including the symmetry factors

\begin{align}
& T_{(a)}^{(3)\,\mu\nu}=64\pi^3 G_N^3 m^4\int \prod_{n=1}^3\frac{d^d{\vec l}_n}{(2\pi)^{d}}\frac{\tau_{(3) \pi\rho, \sigma\tau}^{\mu\nu}(l_1+l_2,q)\tau_{(3)}^{\pi\rho}(-l_1,l_1+l_2)\tau_{(3)}^{\sigma\tau}(-l_3,l_3+l_4)}{(\vec{l}_1)^{^2}(\vec{l}_2)^{^2}(\vec{l}_3)^{^2}(\vec{l}_4)^{^2}(\vec{l}_1+\vec{l}_2)^{^2}(\vec{l}_3+\vec{l}_4)^{^2}}\Bigg\vert_{l_1^0=l_2^0=l_3^0=0},\cr
& T_{(b)}^{(3)\,\mu\nu}=256 \pi^3 G_N^3 m^4\int \prod_{n=1}^3\frac{d^d{\vec l}_n}{(2\pi)^{d}}\frac{\tau_{(3) \sigma\tau,00}^{\mu\nu}(l_1+q,q)\tau_{(3)}^{\pi\rho}(-l_3,l_3+l_4)\tau_{(3) 00,\pi\rho}^{\sigma\tau}(-l_2,l_1+q)}{(\vec{l}_1)^{^2}(\vec{l}_2)^{^2}(\vec{l}_3)^{^2}(\vec{l}_4)^{^2}(\vec{l}_1+\vec{q})^{^2}(\vec{l}_3+\vec{l}_4)^{^2}}\Bigg\vert_{l_1^0=l_2^0=l_3^0=0},\cr
& T_{(c)}^{(3)\,\mu\nu}=-\frac{512 \pi^3 G_N^3m^4}{3}\int \prod_{n=1}^3\frac{d^d{\vec l}_n}{(2\pi)^{d}}\frac{\tau_{(3) \alpha\beta,00}^{\mu\nu}(l_1+q,q)\tau_{(4)00,00,00}^{\alpha\beta}(l_1+q,l_2,l_3,l_4)}{(\vec{l}_1)^{^2}(\vec{l}_2)^{^2}(\vec{l}_3)^{^2}(\vec{l}_4)^{^2}(\vec{l}_1+\vec{q})^{^2}}\Bigg\vert_{l_1^0=l_2^0=l_3^0=0},\cr
& T_{(d)}^{(3)\,\mu\nu}=-256\pi^3 G_N^3m^4\int \prod_{n=1}^3\frac{d^d{\vec l}_n}{(2\pi)^{d}}\frac{\tau_{(3)}^{\gamma\delta}(-l_3,l_3+l_4)\tau_{(4)\gamma\delta,00,00}^{\mu\nu}(q,l_1,l_2,l_3+l_4)}{(\vec{l}_1)^{^2}(\vec{l}_2)^{^2}(\vec{l}_3)^{^2}(\vec{l}_4)^{^2}(\vec{l}_3+\vec{l}_4)^{^2}}\Bigg\vert_{l_1^0=l_2^0=l_3^0=0},\cr
&T_{(e)}^{(3)\,\mu\nu}=\frac{256\pi^3G_N^3m^4}{3}\int \prod_{n=1}^3\frac{d^d{\vec l}_n}{(2\pi)^{d}}\frac{\tau_{(5) 00,00,00,00}^{\mu\nu}(q,l_1,l_2,l_3,l_4)}{(\vec{l_1})^{^2}(\vec{l_2})^{^2}(\vec{l_3})^{^2}(\vec{l_4})^{^2}}\Bigg\vert_{l_1^0=l_2^0=l_3^0=0},
\end{align}
with the five-graviton vertex contribution
\begin{align}\label{e:tau5}
 & \tau^{\mu\nu}_{(5)\, 00,00,00,00}(k_1,k_2,k_3,k_4,k_5):=\tilde
  \tau^{\mu\nu}_{(5)\, \alpha\beta, \gamma\delta, \epsilon\eta, \kappa\lambda}(k_1,k_2,k_3,k_4,k_5) \mathcal P^{\alpha\beta}_{00}
  \mathcal P^{\gamma\delta}_{00} \mathcal
  P^{\epsilon\eta}_{00}P^{\kappa\lambda}_{00} \nn \\
  &=\frac{1}{4(d-1)^3}\Bigg(
 4\delta_{\mu}^0\delta_{\nu}^0\bigg(4(2d^3-18d^2+57d-61) k_1^2+(d-2)(8d^2-47d+79)\sum_{i=2}^5 k_i^2\bigg)\nn \\
 &-(d-2)\eta_{\mu\nu}\bigg((29d^2-191d+362)  k_1^2+(7d^2-61d+142)\sum_{i=2}^5 k_i^2 \bigg)\nn \\
&+2(d-2)\bigg((11d^2-73d+150) k_{1\mu}k_{1\nu}+(7d^2-53d+102)(k_{2\mu}k_{2\nu}+k_{3\mu}k_{3\nu}+k_{4\mu}k_{4\nu}+k_{5\mu}k_{5\nu})\bigg)\Bigg)\,.
\end{align}
where the  vertex $\tau^{\mu\nu}_{(5)\, \alpha\beta, \gamma\delta,
  \epsilon\eta, \kappa\lambda}(k_1,k_2,k_3,k_4,k_5)$ 
 has been derived using the results of~\cite{Prinz:2020nru}.

The integral reduction is done using the {\tt LiteRed}
code~\cite{Lee:2012cn,Lee:2013mka} in $d$ dimensions.
In agreement with  the
general analysis of section~\ref{sec:mast-integr-class}, 
we find that the classical contribution is proportional to the single master integral 

\begin{equation}\label{e:J4master}
  J_{(3)}(\vec q^2)=\int\frac{d^d{\vec l}_1d^d{\vec l}_2d^d{\vec l}_3}{(2\pi)^{3d}}\frac{\vec{q}^2}{\vec{l_1}^2\vec{l_2}^2\vec{l_3}^2(\vec{l_1}+\vec{l_2}+\vec{l_3}+\vec{q})^2}\,.
\end{equation}

\subsubsection{The $\mu=\nu=0$ component}

\begin{align}
&T_{(a)}^{(3)\,00}=-\frac{32 \pi^3 G_N^3m^4}{3}\frac{3d^5-169d^4+1378d^3-4592d^2+7256d-4752}{(d-4)^2(d-1)^3} J_{(3)}(\vec q^2),\cr
&T_{(b)}^{(3)\,00}=-\frac{128 \pi^3 G_N^3m^4}{3}\frac{68d^6-1003d^5+6211d^4-20820d^3+40020d^2-41584d+17824}{(d-4)(d-3)(3d-4)(d-1)^3} J_{(3)}(\vec q^2),\cr
&T_{(c)}^{(3)\,00}=\frac{64 \pi^3 G_N^3m^4}{3}\frac{37d^5-502d^4+2731d^3-7486d^2+10164d-5256}{(d-3)(3d-4)(d-1)^3}  J_{(3)}(\vec q^2),\cr
&T_{(d)}^{(3)\,00}=\frac{32 \pi^3 G_N^3m^4}{3}\frac{53d^4-615d^3+2690d^2-5572d+4840}{(d-4)(d-1)^3}J_{(3)}(\vec q^2),\cr
&T_{(e)}^{(3)\,00}=64 \pi^3 G_N^3m^4\frac{(6-d)(d^2-7d+14)}{(d-1)^3}J_{(3)}(\vec q^2).
\end{align}

\subsubsection{Contraction with $\eta_{\mu\nu}$}

\begin{align}
&T_{(a)}^{(3)\,\mu\nu}\eta_{\mu\nu}=\frac{32 \pi^3 G_N^3m^4}{3}\frac{85d^6-1126d^5+6307d^4-19114d^3+32944d^2-30472d+11952}{(d-4)^2(d-1)^3} J_{(3)}(\vec q^2),\cr
&T_{(b)}^{(3)\,\mu\nu}\eta_{\mu\nu}=\frac{128 \pi^3 G_N^3m^4}{3}\frac{168d^6-2231d^5+12319d^4-35796d^3+57396d^2-48304d+16736}{(d-4)(3d-4)(d-1)^3}J_{(3)}(\vec q^2),\cr
&T_{(c)}^{(3)\,\mu\nu}\eta_{\mu\nu}=-\frac{64 \pi^3 G_N^3m^4}{3}\frac{147d^6-1801d^5+8727d^4-21555d^3+28942d^2-20148d+5688}{(3d-4)(d-1)^4} J_{(3)}(\vec q^2),\cr
&T_{(d)}^{(3)\,\mu\nu}\eta_{\mu\nu}=-\frac{32 \pi^3 G_N^3m^4}{3} \frac{179 d^5 - 2146 d^4 + 10305 d^3 - 24614 d^2 + 28972 d - 13704}{(d-4)(d-1)^3} J_{(3)}(\vec q^2),\cr
 &T_{(e)}^{(3)\,\mu\nu}\eta_{\mu\nu}=\frac{64 \pi^3 G_N^3m^4}{3}\frac{29d^4-274d^3+973d^2-1484d+852}{(d-1)^3}J_{(3)}(\vec q^2).
\end{align}
\subsubsection{The classical three-loop contribution to the stress-tensor}
Summing up all the contributions we get for  the three-loop
stress-tensor 
\begin{equation}\label{e:Tstaticthreeloop}
\langle T^{(3)}_{\mu\nu}\rangle=\pi^3 G_N^3 m^4\Big(c^{(3)}_1(d)\delta_{\mu}^0\delta_{\nu}^0
+c^{(3)}_2(d)\big({q_{\mu}q_{\nu}\over q^2} -\eta_{\mu\nu}\big) \Big)\,  J_{(3)}(q^2)\,,
\end{equation}
with the master integral
\begin{equation}
     J_{(3)}(q^2)=\frac{ \Gamma \left(8-3 d\over 2\right) \Gamma
   \left(\frac{d-2}{2}\right)^4}{8^d\pi^{3d\over2}\Gamma (2 (d-2))}\,
 |\vec q|^{3(d-2)}\,,
\end{equation}
and the three-loop coefficients are given by
\begin{align}
c^{(3)}_1(d)     &=-\frac{64}{3
                    (d-3)(d-4)^2(d-1)^4}\times\Big(56d^7-889d^6+5868d^5\cr
                    &-20907d^4+43434d^3-52498d^2+33888d-8760\Big),\cr
          c^{(3)}_2(d)          &=-\frac{64}{3
                                  (d-3)(d-4)^2(d-1)^4}\times\Big(45d^7-670d^6+4167d^5\cr
                                  &-14016d^4+27430d^3-30916d^2+18104d-3952\Big).
\end{align}
Using the relations~(\ref{e:h0amp}) we obtained
the three-loop contribution to the metric from the classical
stress-tensor in~(\ref{e:T3loopDiv})  (using the notation for $\rho$ in~\eqref{e:rhodef})
\begin{align}\label{e:hthreeloopdiv}
  h_{0}^{(4)}(r,d)&=\frac{16(d-2)^3(14d^3-85d^2+165d-106)}{3(d-3)(d-4)(d-1)^4}\rho(r,d)^4,\nn \\
 h_{1}^{(4)}(r,d)&=-\frac{8(39 d^7 - 
  691 d^6 + 5155 d^5 - 21077 d^4+ 51216 d^3- 74346 d^2+ 60168 d-21208     )}{3(d-3)(d-4)^2(d-1)^4(3d-8)}\rho(r,d)^4,\nn \\
   h_2^{(4)}(r,d) &=\frac{16(d-2)^2( 45 d^6- 580 d^5  + 3007 d^4- 8002 d^3+ 11426 d^2- 8064 d+1976    )}{3(d-3)(d-4)^2(d-1)^4(3d-8)}\rho(r,d)^4.
\end{align}

%%%%%%%%%%%%%%%%%%%%%%%%%%%%%%%%%%%%%%%%%%%%%%%%%%%%%%%%%%%%%%%%%
\section{Non-minimal couplings and renormalised metric}\label{sec:nonmin}

The stress-tensor  and the metric components have ultraviolet
divergences.  
These divergences can be removed by   the addition of the non-minimal
couplings made from the powers of the covariant derivative  $\nabla_\mu$ acting
on a single power of the Riemann tensor and its contractions.
The Bianchi identity on the Riemann tensor $\nabla_{\mu}
R_{\nu\rho\sigma\lambda}+\nabla_{\nu}
R_{\rho\mu\sigma\lambda}+\nabla_{\rho}
R_{\mu\nu\sigma\lambda}=0$, implies that 
\begin{equation}
  \nabla_\mu  R^\mu{}_{\rho\sigma\lambda}=\nabla_{\sigma}
  R_{\rho\lambda}- \nabla_{\lambda} R_{\rho\sigma}, \qquad \nabla_\mu
  R^\mu{}_\nu=\frac12 \nabla_\nu R\,.
\end{equation}
The counter-terms are powers of covariant derivative acting on a
single power of the Ricci tensor and Ricci scalar. 
Therefore the counter-terms are given by  the following non-minimal couplings
\begin{multline}\label{e:Sctn}
  \delta^{(n)}S^{\rm ct.}  = (G_N m)^{2n\over d-2}  \int d^{d+1}x \sqrt{-g}    \Big(\alpha^{(n)}(d) 
  (\nabla^2)^{n-1} R \partial_\mu\phi \partial^\mu\phi \cr+
\left(\beta^{(n)}_0(d) \nabla_\mu\nabla_\nu  (\nabla^2)^{n-2} R+
  \beta^{(n)}_1(d)  (\nabla^2)^{n-1}  R_{\mu\nu} \right) \partial^\mu \phi \partial^\nu\phi\Big).
\end{multline}
where $\alpha^{(n)}(d)$, $\beta_0^{(n)}(d)$ and  $\beta_1^{(n)}(d)$ are dimensionless coefficients
depending on the space-time dimension.  The power of $G_Nm$ is determined by dimensional analysis, and give the
correct order of $G_N$ in all dimensions.
The first non-minimal coupling with $n=1$ is given by
\begin{equation}\label{e:Sctn1}
 \delta^{(1)}   S^{\rm ct.}= (G_N m)^{2\over d-2}\int d^{d+1}x \sqrt{-g} \,
 \left( \alpha^{(1)}(d)  R \partial_\mu\phi \partial^\mu\phi +\beta^{(1)}(d) R^{\mu\nu} \partial_\mu \phi \partial_\nu \phi\right)\,.
\end{equation}
This non-minimal coupling has been introduced in~\cite{Goldberger:2004jt} in four
dimensions and~\cite{Jakobsen:2020ksu} in five
dimensions. We will see that up to three-loop order the renormalisation of the static metric component
only require the counter-term $\alpha^{(1)}(d) 
R\partial_\mu\phi\partial^\mu \phi$, whereas both
couplings are needed for the cancellation of the stress-tensor
divergences. This coupling is
induced by harmonic gauge condition~\cite{Jakobsen:2020ksu,1821624}
and the value of its coefficient depends on the choice of gauge. In
our gauge, the de Donder gauge, this corresponds to $\alpha=0$ in the
work of~\cite{Jakobsen:2020ksu} and $\xi=\frac14$ in the work
of~\cite{1821624}. Since we are working in fixed gauge we will not
discuss further the gauge dependence of the higher-order non-minimal coupling
coefficients, but we expect that the gauge dependence of these
coefficients will be an extension of the discussion in~\cite[app.~B]{Jakobsen:2020ksu}.

The power of the Newton constant in~(\ref{e:Sctn1}) is an integer only
in four dimensions with $d=3$ and five dimensions $d=4$. Therefore
this counter-term will not appear in dimensions $D\geq 6$.

In four dimensions, from five-loop order, or the sixth post-Minkowskian order $O(G_N^6)$,
one expects that higher derivative non-minimal couplings will be needed
to get  finite stress-tensor components. 
In dimensions five and six, the higher-derivative non-minimal couplings arise at lower loop order.

In five dimensions one needs to consider higher-derivative non-minimal
couplings $\delta^{(n)}S^{\rm ct.}$ with $n\geq2$ for removing the divergences in
the stress-tensor.
The non-minimal coupling at this order is then given by 
\begin{multline}\label{e:Sctn2}
  \delta^{(2)} S^{ct.}= (G_Nm)^{4\over d-2} \int d^{d+1}x\sqrt{-g} \Big(
    \alpha^{(2)}(d) \Box R \partial_\mu\phi \partial^\mu\phi
   \cr +    \left(\beta^{(2)}_0(d)  \nabla_\mu\nabla_\nu R+ \beta^{(2)}_1(d) \Box R_{\mu\nu}\right)
    \partial^\mu \phi\partial^\nu\phi \Big)\,.
  \end{multline}
 We will need the non-minimal coupling
\begin{multline}\label{e:Sctn3}
  \delta^{(3)} S^{ct.}= (G_Nm)^{6\over d-2} \int d^{d+1}x\sqrt{-g} \Big(
    \alpha^{(3)}(d) (\nabla^2)^2 R \partial_\mu\phi \partial^\mu\phi
   \cr +    \left(\beta^{(3)}_0(d)  \nabla_\mu\nabla_\nu \nabla^2 R+ \beta^{(3)}_1(d) (\nabla^2)^2 R_{\mu\nu}\right)
    \partial^\mu \phi\partial^\nu\phi \Big)\,,
  \end{multline}
   for removing the two-loop divergence in the
 stress-tensor in six ($d=5$) dimensions and the three-loop divergence
 in five ($d=4$) dimensions. In five dimensions ($d=4$) the metric, up to $G_N^4$, is renormalised using only the
$n=1$ and the metric is finite to all order in six dimensions ($d=5$).

The higher-order non-minimal couplings
$\delta^{(n)}S^{\rm ct.}$ with $n\geq2$ will not contribute to the
classical limit when inserted into graphs with loops, because they
contribute to higher powers in the momentum transfer $\vec q$,
and are sub-leading with respect to the classical contributions.
Their tree-level insertions will contribute to the renormalisation of
the stress-tensor but thanks to the properties of the Fourier
transform they will not contribute to the metric components.

%------------------------------------------------------------------
\subsection{Tree-level insertions}\label{sec:tree-level-nonmin}
We give the contribution of the insertions of the
non-minimal counter-terms with $n=1$ in~(\ref{e:Sctn1}), with $n=2$
in~(\ref{e:Sctn2}) and with $n=3$ in~(\ref{e:Sctn3}) in the tree-level graph.

%------------------------------------------------------------------------
\subsubsection{Insertion of $\delta^{(1)}S^{\rm ct.}$}\label{e:delta1tree}

The insertion of the non-minimal couplings $\delta^{(1)}S^{\rm ct.}$  in~\eqref{e:Sctn1} into the tree-level diagram 
\begin{equation}
\delta^{(1)}\mathcal M^{(0)}(p_1,q)  =
\begin{gathered}\begin{fmffile}{delta1tree}
    \begin{fmfgraph*}(80,100)
\fmfstraight
\fmfleftn{i}{2}
\fmfrightn{o}{1}
\fmf{fermion,tension=2.5,label=$p_1$}{i1,v1}
\fmf{fermion,tension=2.5,label=$p_2$}{v1,i2}
\fmf{dbl_wiggly,tension=1, label.dist=10,label=$q$}{o1,v1}
\fmfv{decor.shape=square,decor.filled=empty, decor.size=5thick,label=1,label.dist=-2}{v1}
\end{fmfgraph*}
\end{fmffile}
\end{gathered},
\end{equation}
leads  to the stress-tensor  contribution in $d+1$ dimensions
\begin{equation}\label{e:T1nonmin}
  \delta^{(1)}  \langle T_{\mu\nu}^{(0)}\rangle= 
 - \vec{q}^{2} (G_Nm)^{2\over d-2}  m \left( -{\beta^{(1)}(d)}\delta_\mu^0\delta_\nu^0+ 2{\alpha^{(1)}(d)}\left(
    {q_\mu q_\nu\over q^2}-\eta_{\mu\nu}\right)\right)\,,
\end{equation}
and  using~\eqref{e:TtohAmplitudedeDonder} this contributes to the metric components
\begin{align}\label{e:hnonmin0}
  \delta^{(1)} h_{0}^{(1)}(r,d)&=0,\\
 \delta^{(1)} h_1^{(1)}(r,d)&={16 \alpha^{(1)}(d)
                                 \Gamma\left(d\over2\right)\over
                                 \pi^{d-2\over2} }\left((G_Nm)^{1\over d-2}\over r\right)^d,\cr
\delta^{(1)} h_{2}^{(1)}(r,d)&=- {32 \alpha^{(1)}(d)
                                 \Gamma\left(d+2\over2\right)\over
                                 \pi^{d-2\over2} }\left((G_Nm)^{1\over d-2}\over r\right)^d\,.
\end{align}
Thanks to the properties of the Fourier transformation (see
appendix~\ref{sec:FT}) only the coefficient $\alpha(d)$ contributes to
static metric perturbation.
%------------------------------------------------------------------------
\subsubsection{Insertion of $\delta^{(2)}S^{\rm
    ct.}$}\label{e:delta2tree}
The insertion of the non-minimal couplings $\delta^{(2)}S^{\rm ct.}$  in~\eqref{e:Sctn2} into the tree-level diagram 
\begin{equation}
\delta^{(2)}\mathcal M^{(0)}(p_1,q)  =
\begin{gathered}\begin{fmffile}{delta2tree}
    \begin{fmfgraph*}(80,100)
\fmfstraight
\fmfleftn{i}{2}
\fmfrightn{o}{1}
\fmf{fermion,tension=2.5,label=$p_1$}{i1,v1}
\fmf{fermion,tension=2.5,label=$p_2$}{v1,i2}
\fmf{dbl_wiggly,tension=1, label.dist=10,label=$q$}{o1,v1}
\fmfv{decor.shape=square,decor.filled=empty, decor.size=5thick,label=2,label.dist=-2}{v1}
\end{fmfgraph*}
\end{fmffile}
\end{gathered},
\end{equation}
leads to the stress-tensor condition in $d+1$ dimensions
\begin{equation}\label{e:T2nonmin}
  \delta^{(2)}  \langle T_{\mu\nu}^{(0)}\rangle= 
 |\vec{q}|^{4} (G_Nm)^{4\over d-2}  m \left( -{\beta_1^{(2)}(d)}\delta_\mu^0\delta_\nu^0+ 2\left(\alpha^{(2)}(d)+\frac12\beta_0^{(2)}(d)\right)\left(
    {q_\mu q_\nu\over q^2}-\eta_{\mu\nu}\right)\right)\,.
\end{equation}
Because of the vanishing of the Fourier transforms
\begin{equation}
\int_{\mathbb R^d} |\vec q|^2 e^{i\vec q\cdot\vec x}    {d^d\vec q\over (2\pi)^d}=0,\qquad
\int_{\mathbb R^d} {q_iq_j\over |\vec q|^{2}}  |\vec q|^2 e^{i\vec q\cdot\vec x}
 {d^d\vec q\over(2\pi)^d}=0\,,
\end{equation}
this extra contribution to the stress-tensor does not affect the
metric components
\begin{align}\label{e:delta2hnonmin0}
  \delta^{(2)} h_{0}^{(1)}(r,d)&=0,\\
 \delta^{(2)} h_1^{(1)}(r,d)&=0,\cr
\delta^{(2)} h_{2}^{(1)}(r,d)&=0\,.
\end{align}

%------------------------------------------------------------------------
\subsubsection{Insertion of $\delta^{(3)}S^{\rm
    ct.}$}\label{e:delta3tree}
The insertion of the non-minimal couplings $\delta^{(3)}S^{\rm ct.}$  in~\eqref{e:Sctn3} into the tree-level diagram 

\begin{equation}
\delta^{(3)}\mathcal M^{(0)}(p_1,q)  =
\begin{gathered}\begin{fmffile}{delta3tree}
    \begin{fmfgraph*}(80,100)
\fmfstraight
\fmfleftn{i}{2}
\fmfrightn{o}{1}
\fmf{fermion,tension=2.5,label=$p_1$}{i1,v1}
\fmf{fermion,tension=2.5,label=$p_2$}{v1,i2}
\fmf{dbl_wiggly,tension=1, label.dist=10,label=$q$}{o1,v1}
\fmfv{decor.shape=square,decor.filled=empty, decor.size=5thick,label=3,label.dist=-2}{v1}
\end{fmfgraph*}
\end{fmffile}
\end{gathered},
\end{equation}
leads to the stress-tensor condition in six  dimensions ($d=5$)
\begin{equation}\label{e:T3nonmin}
  \delta^{(3)}  \langle T_{\mu\nu}^{(0)}\rangle= 
 - |\vec{q}|^{6} (G_Nm)^{6\over d-2}  m \left( -{\beta_1^{(3)}(d)}\delta_\mu^0\delta_\nu^0+ 2\left(\alpha^{(3)}(d)+\frac14\beta_0^{(3)}(d)\right)\left(
    {q_\mu q_\nu\over q^2}-\eta_{\mu\nu}\right)\right)\,.
\end{equation}
Because of the vanishing of the Fourier transforms
\begin{equation}
\int_{\mathbb R^d} |\vec q|^4 e^{i\vec q\cdot\vec x}    {d^d\vec q\over (2\pi)^d}=0,\qquad
\int_{\mathbb R^d} {q_iq_j\over |\vec q|^{2}}  |\vec q|^4 e^{i\vec q\cdot\vec x}
 {d^d\vec q\over(2\pi)^d}=0\,,
\end{equation}
this extra contribution to the stress-tensor does not affect the
metric components
\begin{align}\label{e:delta3hnonmin0}
  \delta^{(2)} h_{0}^{(1)}(r,d)&=0,\\
 \delta^{(2)} h_1^{(1)}(r,d)&=0,\cr
\delta^{(2)} h_{2}^{(1)}(r,d)&=0\,.
\end{align}
%------------------------------------------------------------------
\subsection{One-loop insertions}
\label{sec:one-loop-nonmin}
We give the contribution of the insertions of the
counter-terms~(\ref{e:Sctn}) with $n=1$ in~(\ref{e:Sctn1}) in the one-loop graph.
%------------------------------------------------------------------------
\subsubsection{Insertion of $\delta^{(1)}S^{\rm ct.}$}\label{e:delta1oneloop}

The insertion of the non-minimal coupling in~(\ref{e:Sctn1}) in
the one-loop graph 
\begin{equation}
\delta^{(1)} \mathcal M^{(1)}(p_1,q)=
\begin{gathered}
\begin{fmffile}{delta1oneloop1}
  \begin{fmfgraph*}(100,100)
      \fmfleftn{i}{2}
      \fmfrightn{o}{1}
\fmf{fermion,label=$p_1$}{i1,v1}
\fmf{fermion,label=$p_2$}{v2,i2}
\fmf{dbl_wiggly,label=$q$}{o1,v3}
\fmf{plain,tension=.1}{v1,v2}
\fmf{dbl_wiggly,tension=.3}{v3,v1}
\fmf{dbl_wiggly,tension=.3}{v3,v2}
\fmfv{decor.shape=square,decor.filled=empty, decor.size=5thick,label=1,label.dist=-3}{v1}
\end{fmfgraph*}
\end{fmffile}
\end{gathered}
+
\begin{gathered}
\begin{fmffile}{delta1oneloop2}
  \begin{fmfgraph*}(100,100)
      \fmfleftn{i}{2}
      \fmfrightn{o}{1}
\fmf{fermion,label=$p_1$}{i1,v1}
\fmf{fermion,label=$p_2$}{v2,i2}
\fmf{dbl_wiggly,label=$q$}{o1,v3}
\fmf{plain,tension=.1}{v1,v2}
\fmf{dbl_wiggly,tension=.3}{v3,v1}
\fmf{dbl_wiggly,tension=.3}{v3,v2}
\fmfv{decor.shape=square,decor.filled=empty, decor.size=5thick,label=1,label.dist=-3}{v2}
\end{fmfgraph*}
\end{fmffile}
\end{gathered},
\end{equation}
leads to the stress-tensor contribution 
\begin{align}\label{e:T3loopct}
\delta^{(1)}\langle T_{\mu\nu}^{(1)}\rangle&=32i\alpha^{(1)}(d)  \pi (G_Nm)^{d\over d-2} m^2\cr
           &\times\int\frac{d^d{\vec l}}{(2\pi)^d}\frac{{\tau}^{\mu\nu}_{10\ \alpha\beta,\gamma\delta}(l,q)\mathcal{P}^{\alpha\beta}_{00}l^{\gamma}l^{\delta}}{l^2(l+q)^2}\Big[\frac{1}{(l+p_1)^2-m^2+i\epsilon}+\frac{1}{(l-p_2)^2-m^2+i\epsilon}\Big] \cr
&=8\pi\alpha^{(1)}(d) (G_Nm)^{d\over d-2}m\vec{q}^2\frac{d-2}{(d-1)^2}\Big(d\delta^0_{\mu}\delta^0_{\nu}+\frac{q_{\mu}q_{\nu}}{q^2}-\eta_{\mu\nu}\Big) J_{(1)}(\vec{q}^2)\,.
\end{align}
where we used that
\begin{equation}
  \eta_{\mu\nu}{\tau}^{\mu\nu}_{10\ 00,\gamma\delta}(l,q)l^{\gamma}l^{\delta}=\frac{\vec{l_1}^2}{2}\bigg(\vec{q}^2+(\vec{l_1}+\vec{q})^2-\vec{l_1}^2\bigg),
\end{equation}
and
\begin{multline}
\delta_{\mu}^0\delta_{\nu}^0 {\tau}^{\mu\nu}_{10\ 00,\gamma\delta}(l,q)l^{\gamma}l^{\delta}=
\frac{1}{2(d-1)}\bigg((d-2)\vec{q}^4+(d-2)(\vec{l_1}+\vec{q})^2\big((\vec{l_1}+\vec{q})^2-2\vec{q}^2\big)-\vec{l_1}^4\cr-(d-3)\vec{l_1}^2\big((\vec{l_1}+\vec{q})^2+\vec{q}^2\big)\bigg)\,.
\end{multline}
Using the Fourier transforms
\begin{align}
\int_{\mathbb R^d}  J_{(1)}(\vec q^2) e^{i\vec
    q\cdot \vec x} {d^d\vec q\over (2\pi)^d}
  &=-{\Gamma\left(d\over2\right)^2\over 2 \pi^d r^{2(d-1)}},\cr
  \int_{\mathbb R^d} {q_iq_j\over \vec q^2} J_{(1)}(\vec q^2) e^{i\vec
    q\cdot \vec x} {d^d\vec q\over (2\pi)^d}
  &={\Gamma\left(d-2\over2\right) \Gamma\left(d\over2\right)\over
    4\pi^d  r^{2(d-1)}} \left(
   \delta_{ij}-2 (d-1) {x_ix_j\over r}\right)\,.
\end{align}
and  the relation between the stress-tensor and the metric components
in~\eqref{e:TtohAmplitudedeDonder} we obtain the following contribution to the metric components
\begin{align}\label{e:hnonmin1}
  \delta^{(1)} h_{0}^{(2)}(r,d)&=64\alpha^{(1)}(d)\frac{(d-2)\Gamma({d\over2})^2}{(d-1)\pi^{d-2}}\left((G_Nm)^{1\over d-2}\over r\right)^{2(d-1)},\cr
 \delta^{(1)} h_1^{(2)}(r,d)&=-64\alpha^{(1)}(d)\frac{\Gamma({d\over2})^2}{(d-1)\pi^{d-2}}\left((G_Nm)^{1\over d-2}\over r\right)^{2(d-1)},\cr
  \delta^{(1)} h_{2}^{(2)}(r,d)&=128\alpha^{(1)}(d)\frac{\Gamma({d\over2})^2}{(d-1)\pi^{d-2}}\left((G_Nm)^{1\over d-2}\over r\right)^{2(d-1)}\,.
\end{align}

%--------------------------------------------------------------------------
\subsubsection{Two insertions of $\delta^{(1)}S^{\rm ct.}$}\label{e:delta1delta1oneloop}

Two insertions of the non-minimal coupling $\delta^{(1)}S^{\rm ct.}$
in~\eqref{e:Sctn1} in the one-loop graph
\begin{equation}
(\delta^{(1)} )^2\mathcal M^{(1)}(p_1,q)=
\begin{gathered}
\begin{fmffile}{delta1delta1oneloop1}
  \begin{fmfgraph*}(100,100)
      \fmfleftn{i}{2}
      \fmfrightn{o}{1}
\fmf{fermion,label=$p_1$}{i1,v1}
\fmf{fermion,label=$p_2$}{v2,i2}
\fmf{dbl_wiggly,label=$q$}{o1,v3}
\fmf{plain,tension=.1}{v1,v2}
\fmf{dbl_wiggly,tension=.3}{v3,v1}
\fmf{dbl_wiggly,tension=.3}{v3,v2}
\fmfv{decor.shape=square,decor.filled=empty,
  decor.size=5thick,label=1,label.dist=-3}{v1}
\fmfv{decor.shape=square,decor.filled=empty, decor.size=5thick,label=1,label.dist=-3}{v2}
\end{fmfgraph*}
\end{fmffile}
\end{gathered},
\end{equation}
leads to the stress-tensor contribution
\begin{equation}\label{abc}
(\delta^1)^2\langle T_{\mu\nu}^{(1)}\rangle=\frac{2(\alpha^{(1)}(d))^2(G_Nm)^{d+2\over{d-2}}\pi m \vec{q}^4}{d-1}\Bigg(\delta_{\mu}^0\delta_{\nu}^0-(d-2)\big({q_{\mu}q_{\nu}\over q^2}-\eta_{\mu\nu}\big)\Bigg)\, J_{(1)}(\vec{q}^2),
\end{equation}
and the metric contributions

\begin{align}
&(\delta^1)^2 h^{(2)}_0(r,d)=0,\cr
&(\delta^1)^2 h^{(2)}_1(r,d)={64 (\alpha^{(1)}(d))^2 \over \pi^{d-2}}\Gamma\left({d\over2}\right)^2\left((G_Nm)^{1\over d-2}\over r\right)^{2d},\cr
&(\delta^1)^2 h^{(2)}_2={64 d(d-2) (\alpha^{(1)}(d))^2 \over \pi^{d-2}}\Gamma\left({d\over2}\right)^2\left((G_Nm)^{1\over d-2}\over r\right)^{2d}.
\end{align}
%-------------------------------------------------------------------------
\subsection{Two-loop insertions}

{\centering
  \begin{table}
    \begin{tabular}{ccc}
    \begin{fmffile}{delta1twolooptriangle4pt12a}
 \begin{fmfgraph*}(100,50)
\fmfstraight
      \fmfleftn{i}{2}
      \fmfrightn{o}{1}
      \fmf{plain,tension=10}{i1,v1,v4,v2,i2}
\fmffreeze
\fmf{dbl_wiggly,tension=3}{v3,v5,v2}
\fmf{dbl_wiggly}{v3,v1}
\fmf{dbl_wiggly,tension=.1}{v5,v4}
\fmf{dbl_wiggly,tension=5}{v3,o1}
\fmfv{decor.shape=square,decor.filled=empty, decor.size=5thick,label=1,label.dist=0}{v1}
\end{fmfgraph*}
\end{fmffile},&
\begin{fmffile}{delta1twolooptriangle4pt12b}
 \begin{fmfgraph*}(100,50)
\fmfstraight
      \fmfleftn{i}{2}
      \fmfrightn{o}{1}
\fmf{plain,tension=10}{i1,v1,v4,v2,i2}
\fmffreeze
\fmf{dbl_wiggly,tension=3}{v3,v5,v2}
\fmf{dbl_wiggly}{v3,v1}
\fmf{dbl_wiggly,tension=.1}{v5,v4}
\fmf{dbl_wiggly,tension=5}{v3,o1}
\fmfv{decor.shape=square,decor.filled=empty, decor.size=5thick,label=1,label.dist=-2}{v4}
\end{fmfgraph*}
\end{fmffile},&
\begin{fmffile}{delta1twolooptriangle4pt12c}
 \begin{fmfgraph*}(100,50)
\fmfstraight
      \fmfleftn{i}{2}
      \fmfrightn{o}{1}
\fmf{plain,tension=10}{i1,v1,v4,v2,i2}
\fmffreeze
\fmf{dbl_wiggly,tension=3}{v3,v5,v2}
\fmf{dbl_wiggly}{v3,v1}
\fmf{dbl_wiggly,tension=.1}{v5,v4}
\fmf{dbl_wiggly,tension=5}{v3,o1}
\fmfv{decor.shape=square,decor.filled=empty, decor.size=5thick,label=1,label.dist=-2}{v2}
\end{fmfgraph*}
\end{fmffile},\\
\begin{fmffile}{delta1twolooptriangle4pt21a}
  \begin{fmfgraph*}(100,50)
\fmfstraight
      \fmfleftn{i}{2}
      \fmfrightn{o}{1}
\fmf{plain,tension=10}{i1,v1,v4,v2,i2}
\fmffreeze
\fmf{dbl_wiggly,tension=3}{v3,v5,v1}
\fmf{dbl_wiggly}{v3,v2}
\fmf{dbl_wiggly,tension=.1}{v5,v4}
\fmf{dbl_wiggly,tension=5}{o1,v3}
\fmfv{decor.shape=square,decor.filled=empty, decor.size=5thick,label=1,label.dist=-2}{v1}
\end{fmfgraph*}
\end{fmffile},&
               \begin{fmffile}{delta1twolooptriangle4pt21b}
  \begin{fmfgraph*}(100,50)
\fmfstraight
      \fmfleftn{i}{2}
      \fmfrightn{o}{1}
\fmf{plain,tension=10}{i1,v1,v4,v2,i2}
\fmffreeze
\fmf{dbl_wiggly,tension=3}{v3,v5,v1}
\fmf{dbl_wiggly}{v3,v2}
\fmf{dbl_wiggly,tension=.1}{v5,v4}
\fmf{dbl_wiggly,tension=5}{o1,v3}
\fmfv{decor.shape=square,decor.filled=empty, decor.size=5thick,label=1,label.dist=-2}{v4}
\end{fmfgraph*}
\end{fmffile},&
\begin{fmffile}{delta1twolooptriangle4pt21c}
  \begin{fmfgraph*}(100,50)
\fmfstraight
      \fmfleftn{i}{2}
      \fmfrightn{o}{1}
\fmf{plain,tension=10}{i1,v1,v4,v2,i2}
\fmffreeze
\fmf{dbl_wiggly,tension=3}{v3,v5,v1}
\fmf{dbl_wiggly}{v3,v2}
\fmf{dbl_wiggly,tension=.1}{v5,v4}
\fmf{dbl_wiggly,tension=5}{o1,v3}
\fmfv{decor.shape=square,decor.filled=empty, decor.size=5thick,label=1,label.dist=-2}{v2}
\end{fmfgraph*}
\end{fmffile},\\
\begin{fmffile}{delta1twolooptriangle3a}
  \begin{fmfgraph*}(100,50)
\fmfstraight
      \fmfleftn{i}{2}
      \fmfrightn{o}{1}
\fmf{plain}{i1,v1,vph1a,vph1b,vph1c,v4,vph2,v2,i2}
\fmffreeze
\fmf{dbl_wiggly}{v3,v1}
\fmf{dbl_wiggly}{v3,v2}
\fmf{dbl_wiggly,tension=3}{o1,v5,v3}
\fmf{dbl_wiggly,left=.5,tension=.1}{v5,v4}
\fmfv{decor.shape=square,decor.filled=empty, decor.size=5thick,label=1,label.dist=-2}{v1}
\end{fmfgraph*}
\end{fmffile},&
 \begin{fmffile}{delta1twolooptriangle3b}
  \begin{fmfgraph*}(100,50)
\fmfstraight
      \fmfleftn{i}{2}
      \fmfrightn{o}{1}
\fmf{plain}{i1,v1,vph1a,vph1b,vph1c,v4,vph2,v2,i2}
\fmffreeze
\fmf{dbl_wiggly}{v3,v1}
\fmf{dbl_wiggly}{v3,v2}
\fmf{dbl_wiggly,tension=3}{o1,v5,v3}
\fmf{dbl_wiggly,left=.5,tension=.1}{v5,v4}
\fmfv{decor.shape=square,decor.filled=empty, decor.size=5thick,label=1,label.dist=-2}{v4}
\end{fmfgraph*}
\end{fmffile},&
 \begin{fmffile}{delta1twolooptriangle3c}
  \begin{fmfgraph*}(100,50)
\fmfstraight
      \fmfleftn{i}{2}
      \fmfrightn{o}{1}
\fmf{plain}{i1,v1,vph1a,vph1b,vph1c,v4,vph2,v2,i2}
\fmffreeze
\fmf{dbl_wiggly}{v3,v1}
\fmf{dbl_wiggly}{v3,v2}
\fmf{dbl_wiggly,tension=3}{o1,v5,v3}
\fmf{dbl_wiggly,left=.5,tension=.1}{v5,v4}
\fmfv{decor.shape=square,decor.filled=empty, decor.size=5thick,label=1,label.dist=-2}{v2}
\end{fmfgraph*}
\end{fmffile},\\
 \begin{fmffile}{delta1twolooptriangle4pta}
  \begin{fmfgraph*}(100,50)
\fmfstraight
      \fmfleftn{i}{2}
      \fmfrightn{o}{1}
\fmf{plain,tension=10}{i1,ov1,ovph1,ovph2,ov3,ovph3,ovph4,ov2,i2}
\fmffreeze
\fmf{dbl_wiggly,tension=.3}{v3,ov1}
\fmf{dbl_wiggly,tension=.3}{v3,ov2}
\fmf{dbl_wiggly,tension=.3}{v3,ov3}
\fmf{dbl_wiggly,tension=1}{o1,v3}
\fmfv{decor.shape=square,decor.filled=empty, decor.size=5thick,label=1,label.dist=-2}{ov1}
\end{fmfgraph*}
\end{fmffile},&
 \begin{fmffile}{delta1twolooptriangle4ptb}
  \begin{fmfgraph*}(100,50)
\fmfstraight
      \fmfleftn{i}{2}
      \fmfrightn{o}{1}
\fmf{plain,tension=10}{i1,ov1,ovph1,ovph2,ov3,ovph3,ovph4,ov2,i2}
\fmffreeze
\fmf{dbl_wiggly,tension=.3}{v3,ov1}
\fmf{dbl_wiggly,tension=.3}{v3,ov2}
\fmf{dbl_wiggly,tension=.3}{v3,ov3}
\fmf{dbl_wiggly,tension=1}{o1,v3}
\fmfv{decor.shape=square,decor.filled=empty, decor.size=5thick,label=1,label.dist=-2}{ov3}
\end{fmfgraph*}
\end{fmffile},&
 \begin{fmffile}{delta1twolooptriangle4ptc}
  \begin{fmfgraph*}(100,50)
\fmfstraight
      \fmfleftn{i}{2}
      \fmfrightn{o}{1}
\fmf{plain,tension=10}{i1,ov1,ovph1,ovph2,ov3,ovph3,ovph4,ov2,i2}
\fmffreeze
\fmf{dbl_wiggly,tension=.3}{v3,ov1}
\fmf{dbl_wiggly,tension=.3}{v3,ov2}
\fmf{dbl_wiggly,tension=.3}{v3,ov3}
\fmf{dbl_wiggly,tension=1}{o1,v3}
\fmfv{decor.shape=square,decor.filled=empty, decor.size=5thick,label=1,label.dist=-2}{ov2}
\end{fmfgraph*}
\end{fmffile} .              
    \end{tabular}
    \caption{Insertion of the non-minimal coupling in the two-loop graph}
\label{fig:2loopinsertions}
  \end{table}
}

For the insertion of the non-minimal coupling $\delta^{(1)}S^{\rm ct.}$
in~\eqref{e:Sctn1} in the two-loop graph one needs to sum over all the
contributions in table~\ref{fig:2loopinsertions}.
The  classical limit of the sum of all these graphs lead to the
following contribution to the stress-tensor 
\begin{multline}\label{e:delta1T2}
\delta^{(1)}\langle T_{\mu\nu}^{(2)}\rangle =-{128\pi^2(d-2)\alpha^{(1)}(d)\over 3(d-4)(3d-4)(d-1)^2}
  (G_Nm)^{2(d-1)\over d-2}m \vec{q}^2\Bigg((3d^3-19d^2+28d-10)\delta_{\mu}^0\delta_{\nu}^0\cr+ (3d^3-15d^2+18d-4)\big({q_{\mu}q_{\nu}\over q^2}-\eta_{\mu\nu}\big)\Bigg) J_{(2)}(\vec{q}^2),
\end{multline}
which leads to the following contributions to the metric components
\begin{align}
  \delta^{(1)} h_{0}^{(3)}(r,d)&=-{512\alpha^{(1)}(d)\over d-1}\frac{\Gamma({d\over2})^3}{\pi^{{3\over2}(d-2)}}\left((G_Nm)^{1\over d-2}\over r\right)^{3d-4},\\
 \delta^{(1)} h_1^{(3)}(r,d)&=\frac{256 \alpha^{(1)}(d)\left(3 d^3-23 d^2+46 d-28\right)}{(d-4) (d-2) (d-1)^2 (3 d-4)}\frac{\Gamma({d\over2})^3}{\pi^{{3\over2}(d-2)}}\left((G_Nm)^{1\over d-2}\over r\right)^{3d-4},\cr
\nonumber  \delta^{(1)} h_{2}^{(3)}(r,d)&=-\frac{256\alpha^{(1)}(d) \left(3 d^3-15 d^2+18 d-4\right)}{(d-4) (d-2) (d-1)^2}\frac{\Gamma({d\over2})^3}{\pi^{{3\over2}(d-2)}}\left((G_Nm)^{1\over d-2}\over r\right)^{3d-4}.
\end{align}

%---------------------------------------------------------------------
\subsection{The renormalised  metric in four
  dimensions}\label{sec:renorD4}

The metric components have ultraviolet poles in four dimensions from
two-loop order.  We show how the addition of the non-minimal couplings
leads to finite renormalised metric components.

\subsubsection{The two-loop renormalisation}
\label{sec:two-loop-form}

The two-loop metric components in~\eqref{e:htwoloopdiv} have a divergence in
four  dimensions  ($d=3$)
\begin{align}\label{e:htwoloopdivpole4D}
  h_{0}^{(3)}(r,d)& =O(1),\cr
h_1^{(3)}(r,d)&=-{2\over 3(d-3)}\left(G_Nm\over r\right)^3+O(1),\cr
 h_{2}^{(3)}(r,d)&={2\over d-3}\left(G_Nm\over r\right)^3+O(1)\,.
\end{align}
This divergence is cancelled by adding the metric contribution from
the non-minimal coupling in~(\ref{e:hnonmin0})
\begin{equation}
  h_i^{\rm renor.~(3)}(r,d):=    h_i^{(3)}(r,d)+ \delta^{(1)} h_i^{(1)}(r,d),  \qquad i=0,1,2\,
\end{equation}
and setting the $\alpha^{(1)}(d)$ coefficient to be
\begin{equation}\label{e:alpha4d}
  \alpha^{(1)}(d)={1\over 12(d-3)}+a^{(1)}(3)-{\log(2)\over6}+O(d-3)\,.
\end{equation}
The resulting renormalised two-loop metric reads
\begin{align}\label{e:htwolooprenorm}
  h_{0}^{\rm renor.~(3)}(r,d)&= 2\left(G_Nm\over r\right)^3+O(d-3),\cr
h_1^{\rm renor.~(3)}(r,d)&= {4\over3} \left(-{1\over2}+6a^{(1)}(3)+\log\left(r C_E\over G_Nm\right)
                 \right)    \left(G_N m \over r\right)^3 +O(d-3),\cr
 h_{2}^{\rm renor.~(3)}(r,d)&=   4 \left({1\over3}-6a^{(1)}(3)- \log\left(rC_E\over G_Nm\right)\right) \left(G_N m \over r\right)^3+O(d-3)\,.
\end{align}
where we have introduced the following combination of the
Euler-Mascheroni constant~\cite{Lagarias} and $\pi$
\begin{equation}
  \label{e:Cedef}
  C_E:= \sqrt{\pi} e^{ \gamma_E\over2} \,.
\end{equation}
The divergence in the two-loop stress-tensor
in~\eqref{e:Tstatictwoloop} 
\begin{equation}\label{e:Ttwoloopdiv}
     \langle T_{\mu\nu}^{(2)}\rangle= 
  {G_N^2 \vec q^2m^3\over 6(d-3)}   \Bigg(2\delta_\mu^0 \delta_\nu^0
  + \left({q_\mu
        q_\nu\over q^2}-\eta_{\mu\nu}\right)\Bigg)+O(1)\,,
  \end{equation}
 is cancelled by adding the contribution in~\eqref{e:T1nonmin} from
 the non-minimal coupling with the following choice of $\beta^{(1)}(d)$  coefficient
\begin{equation}\label{e:beta4d}
  \beta^{(1)}(d)=-{1\over3(d-3)}+O(1)\,.
\end{equation}
Notice that this computation does not determine the finite part of the
$\alpha^{(1)}(d)$ and $\beta^{(1)}(d)$.  They are free scales in the
logarithms. We will show in section~\ref{sec:matching-amplitude} that
this freedom is totally reabsorbed in the change of coordinate and the
Schwarzschild-Tangherlini metric does not have any ambiguity.

%--------------------------------------------------------------------
\subsubsection{The three-loop renormalisation}

The three-loop metric components in~\eqref{e:hthreeloopdiv} have a
divergence in four dimensions ($d=3$) given by
\begin{align}\label{e:hthreeloopdivexp}
  h_{0}^{(4)}(r,d)&=-{2\over 3(d-3)}\left(G_N m\over r\right)^4+O(1),\cr
 h_{1}^{(4)}(r,d)&={2\over 3(d-3)}\left(G_N m\over r\right)^4+O(1),\cr
   h_2^{(4)}(r,d) &=-{4\over 3(d-3)}\left(G_N m\over r\right)^4+O(1),\,.
\end{align}
Adding to this contribution the~(\ref{e:hnonmin1})  from the insertion
of the non-minimal couplings at one-loop, and using the value of
$\alpha^{(1)}(d)$ determined in~\eqref{e:alpha4d}, we obtain the
renormalised three-loop metric
\begin{align}\label{e:hthreelooprenorm}
  h_{0}^{\rm renorm. (4)}(r)&=\left(-{32\over3}+8a^{(1)}(3)+{4\over3}\log\left(rC_E\over G_Nm \right) \right)\left(G_N m\over r\right)^4+O(d-3),\cr
 h_{1}^{\rm renorm. (4)}(r)&=\left(10-8a^{(1)}(3)-{4\over3}\log\left(rC_E\over G_Nm\right)\right)\left(G_N m\over r\right)^4+O(d-3),\cr
 h_2^{\rm renorm. (4)}(r)&=\left(-{86\over3}+16a^{(1)}(3)+{8\over 3}\log\left(rC_E\over G_Nm\right)\right) \left(G_N m\over r\right)^4+O(d-3)\,.
\end{align}
The classical three-loop contribution to the stress-tensor has an
ultraviolet divergence
\begin{equation}\label{e:T3loopDiv}
     \langle T_{\mu\nu}^{(3)}(\vec q)\rangle=-{\pi
     G_N^3m^4 |\vec q|^{3\over2}\over 48(d-3)}
 \Bigg(3\delta_\mu^0 \delta_\nu^0
 +
\left({q_\mu        q_\nu\over q^2}-\eta_{\mu\nu}\right)\Bigg) +O(1)\,,
\end{equation}
this divergence is cancelled by the addition of the contribution
in~\eqref{e:T3loopct} from the non-minimal coupling and the choice of
$\alpha^{(1)}(d)$ in~\eqref{e:alpha4d}.

%---------------------------------------------------------------------
\subsection{The renormalised metric in five
  dimensions}\label{sec:renorD5}

The metric components have ultraviolet divergences in five dimensions from
one-loop order.  We show how the addition of the non-minimal couplings
leads to finite renormalised metric components.

\subsubsection{The one-loop renormalisation}\label{sec:D5ctoneloop}

The metric components in~(\ref{e:honeloop}) have a divergence in
five dimension $(d=4)$ given by 
\begin{align}\label{e:honeloopdiv}
  h_{0}^{(2)}(r,d)&=O(1),\cr
h_1^{(2)}(r,d)&=-\frac{40 }{9 (d-4)}\left(G_N m\over \pi r^2\right)^2+O(1),\cr
 h_2^{(2)}(r,d)&=\frac{160 }{9(d-4) }\left(G_N m\over \pi r^2\right)^2+O(1) \,.
\end{align}
The divergences in the metric components~\eqref{e:honeloopdiv} are
cancelled for the choice 
\begin{equation}\label{e:alpha5d}
    \alpha^{(1)}(d)= {5\over18\pi (d-4)}+a^{(1)}(5)+O(d-4)\,,
\end{equation}
so that the renormalised metric components
\begin{equation}
  h_i^{\rm renor.~(2)}(r,d):=    h_i^{(2)}(r,d)+ \delta^{(1)} h_i^{(1)}(r,d),  \qquad i=0,1,2\,,
\end{equation}
have a finite  expansion  near $d=4$
\begin{align}\label{e:honelooprenorm5D}
  h_{0}^{\rm renor.~(2)}(r,d)&={32\over9}\left(G_Nm\over \pi r^2\right)^2+O(d-4),\\
h_1^{\rm renor.~(2)}(r,d)&={20\over9} \left({14\over15}+{36 a^{(1)}(5)\pi\over5}+ \log\left(r^2C_E^2\over G_Nm\right)
                 \right)    \left(G_N m \over\pi r^2\right)^2 +O(d-4),\cr
\nonumber h_{2}^{\rm renor  (2)}(r,d)&=-{80\over9}    \left({7\over30}+{36a^{(1)}(5)\pi\over5}+ \log\left(r^2C_E^2\over G_Nm\right)\right) \left(G_N m \over \pi r^2\right)^2+O(d-4)\,.
\end{align}
where $C_E$ is defined in~\eqref{e:Cedef}.

Thanks to the properties of the Fourier transform, only the
coefficient $\alpha^{(1)}(d)$ enters the counter-term contribution to the
metric component.
To determine as well the coefficient $\beta^{(1)}(d)$ in~\eqref{e:Sctn1} one
needs to look at the divergences of the stress-tensor
\begin{equation}\label{afe}
  \langle T_{\mu\nu}^{(1)}\rangle= {G_N m^2 \vec q^2\over
    18\pi (d-4)}   \left(7\delta_\mu^0\delta_\nu^0+10 \left( {q_\mu
        q_\nu\over q^2}-\eta_{\mu\nu}\right)\right)+O(1)
\end{equation}
The cancellation of the pole fixes the pole part of $\beta^{(1)}(d)$ near
five dimensions
\begin{equation}
    \beta^{(1)}(d)=-\frac{7}{18\pi (d-4)}+O(1)\,.
  \end{equation}
 %-----------------------------------------------------------
\subsubsection{The two-loop renormalisation}

The two-loop metric components in~(\ref{e:htwoloopdiv}) have a divergence in
five dimensions ($d=4$)
\begin{align}\label{e:htwoloopdivpole5D}
  h_{0}^{(3)}(r,d)& =-{320\over 27(d-4)}\left(G_Nm\over \pi r^2\right)^3  +O(1),\cr
h_1^{(3)}(r,d)&={160\over 27(d-4)}\left(G_Nm\over \pi r^2\right)^3+O(1),\cr
 h_{2}^{(3)}(r,d)&=-{320\over 27(d-4)}\left(G_Nm\over \pi r^2\right)^3+O(1)\,.
\end{align}
The divergences in the metric components~\eqref{e:htwoloopdiv} are
cancelled for the choice made at one-loop in~\eqref{e:alpha5d},
so that the renormalised metric components
\begin{equation}
  h_i^{\rm renor.~(3)}(r,d):=    h_i^{(3)}(r,d)+ \delta^{(1)} h_i^{(2)}(r,d),  \qquad i=0,1,2\,,
\end{equation}
have a finite  expansion  near $d=4$
\begin{align}\label{e:htwolooprenorm5D}
  h_{0}^{\rm renor.~(3)}(r,d)&= {160\over27}\left({2\over15}+{36 a^{(1)}(5)\pi\over5}+ \log\left(r^2C_E^2\over G_Nm\right)
                 \right)   \left(G_Nm\over \pi r^2\right)^3+O(d-4),\cr
h_1^{\rm renor.~(3)}(r,d)&=-{80\over27} \left({7\over15}+{36 a^{(1)}(5)\pi\over5}+ \log\left(r^2C_E^2\over G_Nm\right)
                 \right)    \left(G_N m \over\pi r^2\right)^3 +O(d-4),\cr
 h_{2}^{\rm renor  (3)}(r,d)&= {160\over27}    \left(-{1\over15}+{36a^{(1)}(5)\pi\over5}+ \log\left(r^2C_E^2\over G_Nm\right)\right) \left(G_N m \over \pi r^2\right)^3+O(d-4)\,.
\end{align}

The two-loop stress-tensor  in~\eqref{e:Tstatictwoloop} is not finite
in $d=4$ as it diverges like
\begin{multline}
    \langle T_{\mu\nu}^{(2)}\rangle= {5 G_N^2 m^3 |\vec
      q|^4\over 162\pi^2 (d-4)^2} \left(4\delta_\mu^0\delta_\nu^0+ {q_\mu
        q_\nu\over q^2}-\eta_{\mu\nu}\right)\cr
    + {5 G^2m^3|\vec q|^4\over 162
    \pi^2 (d-4)} \Big(\left(4 \log \left(\vec q^2\over 4\pi\right)+4 \gamma_E -{183\over20}\right)\delta_\mu^0\delta_\nu^0 \cr+\left(\log \left(\vec q^2\over4\pi\right)+ \gamma_E -{41\over20}\right)\left( {q_\mu
      q_\nu\over q^2}-\eta_{\mu\nu}\right)\Big)+O(1)\,.
\end{multline}
The addition of the counter-term in~\eqref{e:T3loopct}  from the
non-minimal couplings in~(\ref{e:Sctn1}) is not enough for making the
stress-tensor finite in $d=4$
\begin{multline}\label{e:T2deltaT1}
    \langle T_{\mu\nu}^{(2)}\rangle+\delta^{(1)}\langle
    T_{\mu\nu}^{(1)}\rangle= - {5 G_N^2 m^3 |\vec
      q|^4\over 162\pi^2 (d-4)^2} \left(4\delta_\mu^0\delta_\nu^0+ {q_\mu
        q_\nu\over q^2}-\eta_{\mu\nu}\right)\cr
    +{5G^2m^3|\vec q|^4\over 162
    \pi^2 (d-4)} \Bigg(\left(4\log \left( G_Nm\right)-{144\pi a^{(1)}(5)\over5}-{109\over60} \right)\delta_\mu^0\delta_\nu^0 \cr
    +\left(\left(\log \left(G_Nm\right) +{17\over60}-{36\over5} \pi a^{(1)}(5)\right)\left( {q_\mu
      q_\nu\over q^2}-\eta_{\mu\nu}\right)\right)\Bigg)+O(1)\,.
\end{multline}
We need to consider the addition of the  counter-term from the
insertion of $\delta^{(2)}S^{\rm ct.}$ evaluated in
section~\ref{e:delta2tree} with the values of the coefficient near $d=4$
\begin{align}
  \beta_1^{(2)}(d)&= {1\over \pi^2}\left({10\over 81(d-4)^2 }+{109 +
                    1728 \pi a^{(1)}(5)\over 1944(d-4)}
                    +a^{(2)}(5)+O(d-4)\right),\cr
            \alpha^{(2)}(d)+\frac12\beta_0^{(2)}(d)&=-{1\over 2\pi^2}\left({5\over 162(d-4)^2 }+\frac{432 \pi  a^{(1)}(5)-17}{1944(d-4)}+b^{(2)}(5)+O(d-4)\right) \,,        
\end{align}
plugged in~\eqref{e:T2nonmin} cancel the divergences in~\eqref{e:T2deltaT1}
\begin{equation}
  \langle T_{\mu\nu}^{(2)}\rangle+\delta^{(1)}\langle
    T_{\mu\nu}^{(1)}\rangle+\delta^{(2)}\langle T^{(0)}_{\mu\nu}\rangle  =O(1)\,.
\end{equation}

%--------------------------------------------------------------------
\subsubsection{The three-loop renormalisation}

The three-loop metric components in~(\ref{e:hthreeloopdiv}) have a divergence in
five dimensions ($d=4$)
\begin{align}\label{e:hthreeloopdivpole5D}
  h_{0}^{(4)}(r,d)& ={1280\over 27(d-4)}\left(G_Nm\over \pi r^2\right)^4  +O(1),\cr
h_1^{(4)}(r,d)&=\left({400\over 81(d-4)^2}-\frac{20\left(101+120\log\left(r^2C_E^2\right)\right)}{243(d-4)}\right)\left(G_Nm\over \pi r^2\right)^4+O(1),\cr
 h_{2}^{(4)}(r,d)&=\left({3200\over 81(d-4)^2}+\frac{160\left(187-120\log \left(r^2C_E^2\right)\right)}{243}\right)\left(G_Nm\over \pi r^2\right)^4+O(1)\,.
\end{align}
The divergences in the metric components~\eqref{e:hthreeloopdiv} are
cancelled for the choice made at one-loop in~\eqref{e:alpha5d},
so that the renormalised metric components
\begin{equation}
  h_i^{\rm renor.~(4)}(r,d):=    h_i^{(4)}(r,d)+ \delta^{(1)} h_i^{(3)}(r,d)+(\delta^1)^2 h^{(2)}_i(r,d),  \qquad i=0,1,2\,,
\end{equation}
have a finite  expansion  near $d=4$
\begin{align}\label{e:hthreelooprenorm5D}
  h_{0}^{\rm renor.~(4)}(r,d)=& -{128\over 243}\left(23+324 a^{(1)}(5)\pi+45 \log\left(r^2C_E^2\over G_Nm\right)
                 \right)   \left(G_Nm\over \pi r^2\right)^4+O(d-4),\cr
                 h_1^{\rm renor.~(4)}(r,d)=&{100\over81} \Bigg(\left({36 a^{(1)}(5)\pi\over5}+ \log\left(r^2C_E^2\over G_Nm\right)\right)\left({161\over30}+{36\over5} a^{(1)}(5)\pi+\log\left(r^2C_E^2\over G_Nm\right) \right)+
                 \cr
               &  +{7085\over1800}\Bigg)\times  \left(G_N m \over\pi r^2\right)^4 +O(d-4),\cr
 h_{2}^{\rm renor  (4)}(r,d)=&-{800\over81} \Bigg(\left({36 a^{(1)}(5)\pi\over5}+ \log\left(r^2C_E^2\over G_Nm\right)\right)\left({41\over15}-{36\over5} a^{(1)}(5)\pi-\log\left(r^2C_E^2\over G_Nm\right)\right)+\cr
               &  +{2381\over900}\Bigg)  \times  \left(G_N m \over\pi r^2\right)^4 +O(d-4).
\end{align}

The three-loop stress-tensor  in~\eqref{e:Tstaticthreeloop} is not finite
in $d=4$ as it diverges like
\begin{multline}
    \langle T_{\mu\nu}^{(3)}\rangle= {25 G_N^3 m^4 |\vec
      q|^6\over 5832\pi^3 (d-4)^3} \left(-{1\over2}\delta_\mu^0\delta_\nu^0+ {q_\mu
        q_\nu\over q^2}-\eta_{\mu\nu}\right)\cr
    + {25 G_N^3m^4|\vec q|^6\over 3888
    \pi^3 (d-4)^2} \Big(-{1\over2}\left( \log \left(\vec q^2\over 4\pi\right)+ \gamma_E -{41\over6}\right)\delta_\mu^0\delta_\nu^0 \cr+\left(\log \left(\vec q^2\over4\pi\right)+ \gamma_E -{17\over10}\right)\left( {q_\mu
      q_\nu\over q^2}-\eta_{\mu\nu}\right)\Big)\cr
         + {225 G_N^3m^4|\vec q|^6\over 839808
    \pi^3 (d-4)} 
    \Big({1\over2}\left({70939\over450}+\pi^2-18\left(\log \left(\vec q^2\over 4\pi\right)+ \gamma_E -{41\over6}\right)^2\right)\delta_\mu^0\delta_\nu^0 \cr
    +\left({4769\over450}-\pi^2+18\left(\log \left(\vec q^2\over 4\pi\right)+ \gamma_E -{17\over10}\right)^2\right)\left( {q_\mu
      q_\nu\over q^2}-\eta_{\mu\nu}\right)\Big)+O(1)\,.
\end{multline}
The addition of the counter-terms in $(\delta^1)^2\langle
T_{\mu\nu}^{(1)}\rangle$ in~\eqref{abc}, and $\delta^{(1)}\langle
T_{\mu\nu}^{(2)}\rangle$ in~\eqref{e:delta1T2}  from the
non-minimal couplings in~(\ref{e:Sctn1}) is not enough for making the
stress-tensor finite in $d=4$
\begin{multline}
    \langle T_{\mu\nu}^{(3)}\rangle+(\delta^1)^2\langle T_{\mu\nu}^{(1)}\rangle+\delta^{(1)}\langle T_{\mu\nu}^{(2)}\rangle
    = {25 G_N^3 m^4 |\vec
      q|^6\over 5832\pi^3 (d-4)^3} \left(-{1\over2}\delta_\mu^0\delta_\nu^0+ {q_\mu
        q_\nu\over q^2}-\eta_{\mu\nu}\right)\cr
    + {25 G_N^3m^4|\vec q|^6\over 3888
    \pi^3 (d-4)^2} \Bigg(-{1\over2}\left( {25\over12}+{36\over5}  a^{(1)}(5)\pi -\log\left(G_Nm\right)\right)\delta_\mu^0\delta_\nu^0 \cr+\left({1\over60}+{36\over5}  a^{(1)}(5)\pi -\log\left(G_Nm\right)\right)\left( {q_\mu
      q_\nu\over q^2}-\eta_{\mu\nu}\right)\Bigg)\cr
         - {25 G_N^3m^4|\vec q|^6\over 5184
    \pi^3 (d-4)} \times\cr
   \times \Bigg(\left({27487\over48600}+ a^{(1)}(5)\pi \left(1+{288\over25}a^{(1)}(5)\pi\right) -{\log\left(G_Nm\right)\over3}\left({7\over2}+\log\left(G_Nm\right)+{72\over5}a^{(1)}(5)\pi \right)\right)\delta_\mu^0\delta_\nu^0 \cr
   + \left({6749\over16200}- {6a^{(1)}(5)\pi\over25} \left(1+144a^{(1)}(5)\pi\right) -{\log\left(G_Nm\right)}\left({19\over30}+\log\left(G_Nm\right)-{72\over5}a^{(1)}(5)\pi \right)\right)\left( {q_\mu
      q_\nu\over q^2}-\eta_{\mu\nu}\right)\Bigg)\cr
      +O(1)\,.
\end{multline}
We need to consider the addition of the  counter-term from the
insertion of $\delta^{(3)}S^{\rm ct.}$ evaluated in
section~\ref{e:delta3tree} with the values of the coefficient near $d=4$
\begin{align}
  \beta_1^{(3)}(d)&=\frac{25}{11664 \pi ^3
   (d-4)^3}+ \frac{5 (432 \pi  a^{(1)} (5)+125)}{93312 \pi ^3 (d-4)^2}\cr
&+\frac{559872 (\pi a^{(1)}(5))^2+486000 \pi  a^{(1)} (5)+27487}{6718464 \pi ^3 (d-4)}+O(1),\cr
            \alpha^{(3)}(d)+\frac14\beta_0^{(3)}(d)&=\frac{25}{11664 \pi ^3
   (d-4)^3}+\frac{2160 \pi  a^{(1)}(5)+5}{93312 \pi ^3 (d-4)^2}\cr
&+\frac{559872 (\pi  a^{(1)}(5))^2+3888 \pi  a^{(1)}(5)-6749}{6718464 \pi ^3 (d-4)}+O(1)\,,       
\end{align}
plugged in~\eqref{e:T2nonmin} cancel the divergences in~\eqref{e:T2deltaT1}
\begin{equation}
  \langle T_{\mu\nu}^{(2)}\rangle+\delta^{(1)}\langle
    T_{\mu\nu}^{(2)}\rangle+(\delta^{(1)}\langle T_{\mu\nu}^{(2)}\rangle+\delta^{(3)}\langle
    T_{\mu\nu}^{(0)}\rangle =O(1)\,.
\end{equation}

%---------------------------------------------------------------------
\subsection{The renormalised stress-tensor in six
  dimensions}\label{sec:renorD6}

In six dimensions, the metric component are finite to all order in
perturbation but the two-loop
stress-tensor in~\eqref{e:Tstatictwoloop} presents an ultraviolet
divergence in six dimensions ($d=5$)
\begin{equation}
    \langle T_{\mu\nu}^{(2)}\rangle= -{G_N^2 m^3 |\vec
      q|^6\over 40320\pi^2 (d-5)} \left(49\delta_\mu^0\delta_\nu^0+ 15\left({q_\mu
        q_\nu\over q^2}-\eta_{\mu\nu}\right)\right)+O(1)\,,
\end{equation}
which is cancelled by the addition of the insertion of the non-minimal
coupling $\delta^{(3)}S^{\rm ct.}$ at tree-level in~\eqref{e:T3nonmin}
with the choice of the coefficients
\begin{align}
  \alpha^{(3)}(d)+\frac14\beta^{(3)}_0(d)&=-{15\over80640\pi^2(d-5)}+O(1),\cr
 \beta_1^{(3)}(d)&= -{49\over40320\pi^2(d-5)}+O(1)\,.
\end{align}
%%%%%%%%%%%%%%%%%%%%%%%%%%%%%%%%%%%%%%%%%%%%%%%%%%%%%%%%%%%%%%%%%
\section{The Schwarzschild-Tangherlini  metric in de Donder gauge in
  four, five  and six dimensions }\label{sec:deDonder}

The Schwarzschild-Tangherlini~\cite{Tangherlini:1963bw}  space-time metric in $d+1$ dimensions is given by
the Tangherlini solution, using $\rho(r,d)$ defined in~(\ref{e:rhodef}),\footnote{In
  spherical coordinate  the metric reads
  \begin{equation}
    ds^2 = \left(1-{\mu\over r^{d-2}}\right)dt^2-{dr^2\over
      1-{\mu\over r^{d-1}}}-r^2 d\Omega_{d-1}
    \end{equation}
    with $\mu={16\pi G_Nm\over (d-1)\Omega_{d-1}}$ and
    $\Omega_{d-1}={2\pi^{d\over2}\over \Gamma\left(d\over2\right)}$
      is the area of the unit $(d-1)$-sphere.
}
\begin{equation}
  ds^2_{\rm Schw}= \left(1- 4{d-2\over d-1} \rho(r,d) \right)dt^2-
  d\vec x^2-  {4{d-2\over d-1} \rho(r,d) \over 1- 4{d-2\over d-1} \rho(r,d)}{(\vec x\cdot
    d\vec x)^2\over r^2}\,.
\end{equation}

As explained in section~\ref{sec:schw} the amplitude computation
selects the de Donder gauge in~(\ref{e:deDonderGauge}). 
We make the coordinate transformation $(t,\vec x)\to (t, f(r)\vec x)$
so that  the Schwarzschild metric reads 
\begin{equation}\label{e:SchwaF}
ds^2=h_0(r) dt^2-  h_1(r) d\vec x^2-h_2(r) {(\vec           x\cdot
  d\vec x)^2\over r^2},
\end{equation} 
with $r=|\vec x|$ and
\begin{align}
\label{e:hfinitedef} h_0(r)&:=1-4{d-2\over d-1}\, {\rho(r,d)\over f(r)^{d-2}}, \\
h_1(r)&:=  f(r)^2, \cr
\nonumber h_2(r)&:=-f(r)^2-f(r)^{d-2} {(f(r)+r{df(r)\over dr})^2\over
                  f(r)^{d-2}-4{d-2\over d-1}\rho(r,d) }\,.
\end{align} 
The de Donder gauge condition~(\ref{e:deDonderGauge}) then reads
\begin{equation}\label{e:dDf}
 2(d-1)h_2(r)= r {d\over dr}\left(h_0(r)+(d-2) h_1(r)-h_2(r)\right)\,.
\end{equation}
We will be solving the de Donder gauge
condition~(\ref{e:deDonderGauge}) in four dimensions ($d=3$), five
dimensions ($d=4$) and six dimensions ($d=5$), using the post-Minkowskian
expansion
\begin{equation}\label{e:fpostN}
    f(r)= 1+ \sum_{n\geq1} f_n(r) \rho(r,d)^n
  \end{equation}
  with the condition at each order that
  \begin{equation}\label{e:bdy}
    \lim_{r\to +\infty}     f_n(r)/r^n =0\,.
  \end{equation}
%---------------------------------------------------------------------
\subsection{The metric in the de Donder gauge  in four dimensions}\label{sec:soldeDonder}

The de Donder gauge condition~(\ref{e:deDonderGauge})  in $d=3$ reads
\begin{equation}\label{e:dDfd3}
 4h_2(r)= r {d\over dr}\left(h_0(r)+h_1(r)-h_2(r)\right)\,,
\end{equation}
supplemented with the asymptotic boundary condition
\begin{equation}\label{e:finf4d}
  \lim_{r\to\infty} f(r)=1\,.
\end{equation}

This differential equation implies either that $f(r)=C /r$, which does
not satisfy the boundary condition~(\ref{e:finf4d}), or $f(r)$
satisfies the differential equation, with $x=G_N m/r$ 
\begin{multline}\label{e:diffGauge}
x f(x)^3(2x-f(x)) {d^{2}f(x)\over dx^2}+ \left(x  f \left( x \right) 
 \right) ^{2} \left(df(x)\over dx \right) ^{2}\cr
 +2\,  f \left( x \right)^3 (f(x)-3x ){df(x)\over dx}-3\, \left( f \left( x \right)  \right) ^{4}+8\,
 \left( f \left( x \right)  \right) ^{3}x+ \left( f \left( x \right) 
 \right) ^{2}-4\,f \left( x \right) x+4\,{x}^{2}
=0.
\end{multline}
We solve the equation~(\ref{e:diffGauge}) using a series  expansion in
$G_N m$ using~\eqref{e:fpostN} and the boundary condition~\eqref{e:bdy}.
The result to the order $(G_N m)^7$ is given by 
\begin{multline} \label{e:ffinite}
  f(r)=1+{G_Nm\over r}+2\left( G_N m\over r\right)^2+ {2\over3} \log
   \left(\frac{r C_3}{G_N m}\right)\left(G_N m\over
   r\right)^3\cr
 + \left( {2\over3}-{4\over3}
   \log \left(\frac{r C_3}{G_N m}\right)\right)\left(G_N m\over r\right)^4
+ \left(
  -\frac{21}{25}+{32\over15} \log \left(\frac{rC_3}{G_N m}\right)\right)\left(G_N
  m\over r\right)^5\cr
+ \left(\frac{112}{75}-{28\over15} \log\left(r C_3\over G_N
 m\right)\right) \left(G_N
 m\over r\right)^6\cr
+\left(\frac{50023}{34300}
    +\frac{1139 }{2205} \log\left(r C_3\over G_N
 m\right)+ {2\over 7}\log\left(r C_3\over G_N
 m\right)^2\right)\left(G_N
m\over r\right)^7
+O(G_N^8)\,.
\end{multline}
This solution is finite and has $\log(r)$ terms  from the  order
$G_N^3$. The solution has a single constant of integration  $C_3$
associated with the scale of the logarithm. 

\subsubsection{The metric perturbation}
In $d=3$ we derive components of the metric in  perturbation by
plugging the expression for $f(r)$ in~\eqref{e:ffinite}
in~\eqref{e:hfinitedef}.

We obtain for the time component
\begin{multline}\label{e:h0finite}
  h^{\rm dD}_0(r)=1-2\frac{ G_N m}{r}+2\left(\frac{ G_N
      m}{r}\right)^2+2\left(\frac{ G_N m}{r}\right)^3+\left(\frac43 \log
      \left(\frac{r C_3}{G_N m}\right)-6\right) \left(G_N m\over
      r\right)^4\cr
    +\left(-\frac{16}3 \log
      \left(\frac{rC_3}{G_N m}\right)+{10\over3}\right) \left(G_N m\over
      r\right)^5  +\left(\frac{124}{15} \log
      \left(\frac{rC_3}{G_N m}\right)+\frac{424}{75}\right) \left(G_N m\over
      r\right)^6 \cr
    +\Bigg(-\frac{8}{9} \log
      \left(\frac{rC_3}{G_N m}\right)^2+\frac{16}{15} \log
      \left(\frac{rC_3}{G_N m}\right) 
     -\frac{674}{75}\Bigg) \left(G_N m\over
      r\right)^7
    +O (G_N^8),
\end{multline}
and for the spatial components
\begin{multline}\label{e:h1finite}
  h^{\rm dD}_1(r)=1+2\frac{ G_N m}{r}+5\left(\frac{ G_N m}{r}\right)^2+
    \left({4\over3} \log
      \left(\frac{rC_3}{G_N m}\right)+ 4\right)\left(G_N m\over
      r\right)^3\cr
    +
   \left(-\frac43 \log
      \left(\frac{rC_3}{G_N m}\right) +\frac{16}{3}\right)\left(G_N m\over r\right)^4
      +  \left(\frac{64}{15} \log
      \left(\frac{rC_3}{G_N m}\right) -\frac{26}{75}\right)\left(G_N m\over r\right)^5\cr
       +\Bigg(\frac{4}{9} \log
      \left(\frac{rC_3}{G_N m}\right)^2-\frac{24}{5}\log
      \left(\frac{rC_3}{G_N m}\right) 
     +\frac{298}{75}\Bigg) \left(G_N m\over
      r\right)^6+O(G_N ^7),
\end{multline}
and
\begin{multline}\label{e:h2finite}
  h^{\rm dD}_2(r)=-7\left(\frac{ G_N m}{r}\right)^2-\left(4 \log
      \left(\frac{rC_3}{G_N m}\right)+{38\over3}\right)
  \left(G_N m\over r\right)^3
  + \left(\frac83 \log \left(\frac{rC_3}{G_N
          m}\right)-\frac{58}{3}\right)\left(G_N m\over r\right)^4\cr
           -  \left(\frac{16}{3} \log
      \left(\frac{rC_3}{G_N m}\right) -\frac{32}{3}\right)\left(G_N m\over r\right)^5\cr
       +\Bigg(\frac{4}{3} \log
      \left(\frac{rC_3}{G_N m}\right)^2+\frac{508}{45} \log
      \left(\frac{rC_3}{G_N m}\right) 
      +\frac{7378}{225}\Bigg) \left(G_N m\over
      r\right)^6+O(G_N^7).
\end{multline}
Notice the appearance of the $\log(r)^2$ at the sixth
post-Minkowskian order, $G_N^6$, in the spatial components of the metric. This is
one order less than the appearance in the time component. The same
phenomenon happens for the $\log(r)$ contribution which appears 
one order earlier in the spatial component than in the time component.
%---------------------------------------------------------------------
\subsection{The metric in the de Donder gauge in five
  dimensions}\label{sec:soldeDonder5d}

The de Donder gauge condition~(\ref{e:deDonderGauge})  in $d=4$ reads
\begin{equation}\label{e:dDfd4}
 6h_2(r)= r {d\over dr}\left(h_0(r)+2h_1(r)-h_2(r)\right)\,,
\end{equation}
supplemented with the asymptotic boundary condition
\begin{equation}\label{e:finf5d}
  \lim_{r\to\infty} f(r)=1\,.
\end{equation}

This differential equation implies either that $f(r)=C/r$, which does
not satisfy the boundary condition~(\ref{e:finf5d}), or $f(r)$
satisfies the differential equation, setting $x=G_N m/(\pi r^2)$ 
\begin{multline}\label{e:diffGauged4}
xf(x)^5 \left(8x-3 f(x)^2\right) {d^2f(x)\over dx^2}+8f(x)^4x^2 \left(d
  f(x)\over dx\right)^2+f(x)^5 \left(3f(x)^2-16x\right) {df(x)\over
  dx}\cr
-4f(x)^6+(16x+2)f(x)^4-{32\over3}x f(x)^2+{128x^2\over9}=0\,.
\end{multline}
We solve the equation~(\ref{e:diffGauged4}) using a series  expansion in
$G_N m$ using~\eqref{e:fpostN} and the boundary condition~\eqref{e:bdy}.
The result to the order $(G_N m)^7$ is given by 
\begin{multline}\label{e:fsol5d}
  f(r)=1+\frac23 {G_N m\over \pi r^2}+\frac{10}9 \log\left(r^2C_2\over G_Nm\right) \left(G_N m\over \pi
    r^2\right)^2 -\frac{4}{81}  \left(-8+45  \log\left(r^2C_2\over G_Nm\right)\right)\left(G_N m\over \pi r^2\right)^3
 \cr
  +\frac{67+3780  \log\left(r^2C_2\over G_Nm\right)}{972}
\left(G_N m\over \pi r^2\right)^4-\frac{32963+156420  \log\left(r^2C_2\over G_Nm\right)-43200
    \log\left(r^2C_2\over G_Nm\right)^2}{21870} \left(G_N m\over \pi r^2\right)^5 \cr
+\frac{ 409303+1620270  \log\left(r^2C_2\over G_Nm\right)-1087200  \log\left(r^2C_2\over G_Nm\right)^2}{131220}\left(G_N
  m\over \pi r^2\right)^6\cr
-\frac{
   11148022313+37508666370  \log\left(r^2C_2\over G_Nm\right)-64367301600  \log\left(r^2C_2\over G_Nm\right)^2}{2362944150}\left(G_N m\over
     \pi r^2\right)^7\cr
   -\frac{4939200000 
     }{2362944150}\log\left(r^2C_2\over G_Nm\right)^3\left(G_N m\over \pi r^2\right)^7+O(G_N^8).
\end{multline}
Again there is a single constant of integration $C_2$ arising as the
scale of the $\log(r)$ arising from the $G_N^2$ order.

\subsubsection{The metric perturbation}\label{sec:metpert5D}
In $d=4$ we derive components of the metric in  perturbation by
plugging the expression for $f(r)$ in~\eqref{e:fsol5d}
in~\eqref{e:hfinitedef}.

We obtain for the time component
\begin{multline}\label{e:h0finite5D}
  h^{\rm dD}_0(r)=1-{8\over3} \frac{ G_N m}{ \pi  r^2}+\frac{32}{9 }\left(\frac{ G_N m}{ \pi  r^2}\right)^2+\frac{32 \left(-3+5 \log\left(r^2C_2\over G_N m\right)\right) }{27
   }\left(\frac{ G_N m}{ \pi  r^2}\right)^3\cr
  -\frac{640 \left(-2+9 \log\left(r^2C_2\over G_N m\right)\right)} {243 } \left(\frac{ G_N m}{ \pi  r^2}\right)^4+O(G_N^5)\,,
\end{multline}
and for the spatial components
\begin{multline}\label{e:h1finite5D}
  h^{\rm dD}_1(r)=1+\frac{4 }{3 } \frac{ G_N m}{ \pi
    r^2}+\frac{4\left(1+5 \log\left(r^2C_2\over G_N m\right)\right)
   }{9 } \left(\frac{ G_N m}{ \pi  r^2}\right)^2+\frac{\left(64-240 \log\left(r^2C_2\over G_N m\right)\right) }{81
   }\left(\frac{ G_N m}{ \pi  r^2}\right)^3\cr
  +\frac{\left(323+2340 \log\left(r^2C_2\over G_N m\right)+600 \log
      ^2\left(r^2C_2\over G_Nm\right)\right)}{486 }\left(\frac{ G_N m}{ \pi  r^2}\right)^4+O(G_N^5)\,,
\end{multline}
and
\begin{multline}\label{e:h2finite5D}
  h^{\rm dD}_2(r)=\frac{40 \left(1-2 \log\left(r^2C_2\over G_N
        m\right)\right) }{9 } \left(\frac{ G_N m}{ \pi  r^2}\right)^2+\frac{32 \left(-4+5 \log\left(r^2C_2\over G_N m\right)\right) }{27 }\left(\frac{ G_N m}{ \pi  r^2}\right)^3\cr
  +\frac{8 \left(-31-1260 \log\left(r^2C_2\over G_N m\right)+300 \log\left(r^2C_2\over G_N m\right)^2\right) }{243 }\left(\frac{ G_N m}{ \pi  r^2}\right)^4+O(G_N^5)\,.
\end{multline}
%---------------------------------------------------------------------
\subsection{The metric in the de Donder gauge  in six dimensions}\label{sec:soldeDonder6d}

The de Donder gauge condition~(\ref{e:deDonderGauge})  in $d=5$ reads
\begin{equation}\label{e:dDfd5}
 8h_2(r)= r {d\over dr}\left(h_0(r)+3h_1(r)-h_2(r)\right)\,,
\end{equation}
supplemented with the asymptotic boundary condition
\begin{equation}\label{e:finf6d}
  \lim_{r\to\infty} f(r)=1\,.
\end{equation}

This differential equation implies either that $f(r)= C /r$, which does
not satisfy the boundary condition~(\ref{e:finf6d}), or $f(r)$
satisfies the differential equation with $x=G_Nm/(\pi r^3)$
\begin{multline}\label{e:diffGauged5}
xf(x)^7 \left(6x-4 f(x)^3\right) {d^2f(x)\over dx^2}+9f(x)^6x^2 \left(d
  f(x)\over dx\right)^2+f(x)^7 \left({8\over3}f(x)^3-10x\right) {df(x)\over
  dx}\cr
-{5\over3}f(x)^8+f(x)^6+4xf(x)^5-3x f(x)^3+{9x^2\over4}=0\,.
\end{multline}
We solve the equation~(\ref{e:diffGauged5}) using a series  expansion in
$G_N$ using~\eqref{e:fpostN} and the boundary condition~\eqref{e:bdy}.
Asking for an expression with only integer powers of $G_N$, the result to the order $G_N^7$ is given by 
\begin{multline}\label{e:fsol6d}
  f(r)=1+\frac{G_N m}{4 \pi  r^3}-\frac{5 }{8 }\left(\frac{G_N m}{\pi r^3}\right)^2+\frac{2 }{3 }\left(\frac{G_N m}{\pi r^3}\right)^3-\frac{775}{1344}\left(\frac{G_N m}{\pi r^3}\right)^4+\frac{545977 }{537600 }\left(\frac{G_N m}{\pi r^3}\right)^5\cr
  -\frac{15194099 }{10483200 }\left(\frac{G_N m}{\pi r^3}\right)^6+\frac{4421000509
 }{1878589440}\left(\frac{G_N m}{\pi r^3}\right)^7+O(G_N^8)\,.
\end{multline}
The expression is uniquely determined and finite. 

\subsubsection{The metric perturbation}\label{sec:metpert6D}
In $d=5$ we derive components of the metric in  perturbation by
plugging the expression for $f(r)$ in~\eqref{e:fsol6d}
in~\eqref{e:hfinitedef}.

We obtain for the metric  components
\begin{align}
  h^{\rm dD}_0(r)&=1-\frac{3 G_N m}{2 \pi  r^3}+\frac{9}{8 }\left(\frac{G_N m}{ \pi  r^3}\right)^2-\frac{27 }{8 }\left(\frac{G_N m}{ \pi  r^3}\right)^3+\frac{387 }{64 }\left(\frac{G_N m}{ \pi  r^3}\right)^4+O(G_N^5)\,,\cr
  h^{\rm dD}_1(r)&=1+\frac{G_N m}{2 \pi  r^3}-\frac{19 }{16 }\left(\frac{G_N m}{ \pi  r^3}\right)^2+\frac{49 }{48 }\left(\frac{G_N m}{ \pi  r^3}\right)^3-\frac{577}{1344}\left(\frac{G_N m}{ \pi  r^3}\right)^4+O(G_N^5)\,,\cr
\label{e:h2finite6D}
  h^{\rm dD}_2(r)&=\frac{117}{16 }\left(\frac{G_N m}{ \pi  r^3}\right)^2-\frac{45}{16}\left(\frac{G_N m}{ \pi  r^3}\right)^3+\frac{1599 }{112}\left(\frac{G_N m}{ \pi  r^3}\right)^4+O(G_N^5)\,.
\end{align}

%%%%%%%%%%%%%%%%%%%%%%%%%%%%%%%%%%%%%%%%%%%%%%%%%%%%%%%%%%%%%%%
\section{Recovering the Schwarzschild-Tangherlini  metric from  the amplitude computations}
\label{sec:matching-amplitude}

In this section we show  how the amplitude computations match the Schwarzschild-Tangherlini  metric in
four, five and six dimensions in the de Donder gauge of the previous section.

\subsection{The Schwarzschild metric in four dimensions}

\subsubsection{The first post-Minkowskian contribution $O(G_N)$}

Setting $d=3$ in the expressions for the metric perturbation from the tree-level amplitude
 in~\eqref{e:htree} matches the de Donder gauge first post-Minkowskian order
 in four dimension $(d=3)$ in~(\ref{e:h0finite})--(\ref{e:h2finite}).

\subsubsection{The second post-Minkowskian contribution $O(G_N^2)$}

At the order $G_N^2$, setting $d=3$ in the metric perturbation from the one-loop amplitude
 in~\eqref{e:honeloop} matches the metric in  the de Donder gauge 
 in four dimensions $(d=3)$ in~(\ref{e:h0finite})--(\ref{e:h2finite}).

\subsubsection{The third post-Minkowskian contributions $O(G_N^3)$}

At this order the components of the metric in the de Donder gauge in
four  dimensions ($d=3$)
from~(\ref{e:h0finite})---(\ref{e:h2finite}) match  the 
metric components from the renormalised two-loop
amplitude computation in~(\ref{e:htwolooprenorm}) for the value of the constant
of integration  
\begin{equation}\label{e:C3}
\log C_3= \log C_E-{7\over2}+6a^{(1)}(3)\,,
\end{equation}
where $ C_E$ is given in~\eqref{e:Cedef}.

With this identification we recover the results
of~\cite{Goldberger:2004jt} for the renormalisation of the metric
divergences and the coordinate change from the de Donder gauge to the
harmonic gauge from the world-line approach.

Substituting this value of $C_3$ in the solution~(\ref{e:ffinite})
completely determines the solution to the de Donder gauge in four  dimensions
and the coordinate change in~\eqref{e:ffinite} to the Schwarzschild
metric in~(\ref{e:SchwaF}) in four dimensions. The parameter
$a^{(1)}(3)$ is a free parameter, which corresponds to the running
coupling in~\cite{Goldberger:2004jt}.

\subsubsection{The fourth post-Minkowskian contribution $O(G_N^4)$}
At the fourth post-Minkowskian  order, we get again a diverging metric from the amplitude computation.
This finite component metric in the de Donder gauge in four dimensions
$(d=3)$
in~(\ref{e:h0finite})---(\ref{e:h2finite}) using the value of the
constant of integration $C_3$ determined in~(\ref{e:C3}) give 
\begin{align}
  h^{\rm dD (4)}_{0}&= \left(-{32\over3}+8 a^{(1)}(3)+{4\over3} \log
      \left(\frac{rC_E}{G_N m}\right)\right) \left(G_Nm\over r\right)^4,\\
h^{\rm dD(4)}_1&= 
    \left(10-8a^{(1)}(3)-{4\over3} \log
      \left(\frac{rC_E}{G_N m}\right)\right)\left(G_N m\over r\right)^4,\cr
\nonumber h^{\rm dD(4)}_{2}&=   \left(-{86\over 3}+16a^{(1)}(3)+ {8\over3} \log \left(\frac{rC_E}{G_N
          m}\right)\right)
  \left(G_N m\over r\right)^4\,.
\end{align}
This matches exactly the  renormalised metric components from the three-loop
amplitude computation obtained in~(\ref{e:hthreelooprenorm}) with $d=3$.

\subsection{The Schwarzschild-Tangherlini  metric in five dimensions}

\subsubsection{The first post-Minkowskian contribution $O(G_N)$}

Setting $d=4$ in the expressions for the metric perturbation from the tree-level amplitude
 in~\eqref{e:htree} matches the de Donder gauge first post-Minkowskian order
 in 
 five dimensions $(d=4)$  in~(\ref{e:h0finite5D})--(\ref{e:h2finite5D}).

\subsubsection{The second post-Minkowskian contribution $O(G_N^2)$}
The renormalised one-loop computation in~\eqref{e:honelooprenorm5D}
matches the expression at order $O(G_N^2)$  from the de Donder gauge
in~(\ref{e:h0finite5D})---(\ref{e:h2finite5D})   for the choice of the constant of
integration
\begin{equation}\label{e:C2}
\log  C_2={11\over15} +2\log C_E + {36\pi\over 5}  a^{(1)}(5)\,.
\end{equation}
Again there is a free parameter $a^{(1)}(5)$  which can be associated with a running coupling constant.

\subsubsection{The third post-Minkowskian contributions $O(G_N^3)$}
At this  order in perturbation, the  two-loop amplitude computation
had divergences that had to be renormalized to give~(\ref{e:htwolooprenorm5D}).
This  matches exactly the  finite component metric in the de Donder gauge in five dimensions
$(d=4)$ in~(\ref{e:h0finite5D})---(\ref{e:h2finite5D}), using the value of the
constant of integration $C_2$ determined in~(\ref{e:C2}), given by
\begin{align}\label{e:hhthreeloopdD5}
  h^{\rm dD (3)}_{0}&= {160\over27}\left({2\over15}+{36 a^{(1)}(5)\pi\over5}+ \log\left(r^2C_E^2\over G_Nm\right)
                 \right)   \left(G_Nm\over \pi r^2\right)^3+O(d-4),\\
h^{\rm dD(3)}_1&= -{80\over27} \left({7\over15}+{36 a^{(1)}(5)\pi\over5}+ \log\left(r^2C_E^2\over G_Nm\right)
                 \right)    \left(G_N m \over\pi r^2\right)^3 +O(d-4),\cr
\nonumber h^{\rm dD(3)}_{2}&=   {160\over27}    \left(-{1\over15}+{36a^{(1)}(5)\pi\over5}+ \log\left(r^2C_E^2\over G_Nm\right)\right) \left(G_N m \over \pi r^2\right)^3+O(d-4)\,.
\end{align}

\subsubsection{The fourth post-Minkowskian contribution $O(G_N^4)$}
The three-loop amplitude computation diverges and the finite metric
component at the fourth post-Minkowskian  order was obtained after
normalisation in~(\ref{e:hthreelooprenorm5D}).
This matches exactly, the finite component metric in the de Donder gauge in five dimensions
$(d=4)$ in~(\ref{e:h0finite5D})---(\ref{e:h2finite5D}), using the value of the
constant of integration $C_2$ determined in~(\ref{e:C2}), given by
\begin{align}\label{e:hhthreeloopdD}
  h^{\rm dD (4)}_{0}&=- {128\over 243}\left(23+324 a^{(1)}(5)\pi+ 45\log\left(r^2C_E^2\over G_Nm\right)
                 \right)   \left(G_Nm\over \pi r^2\right)^4+O(d-4),\cr
h^{\rm dD(4)}_1&= \Bigg({7085+69552 \pi
                 a^{(1)}(5)+93312(\pi a^{(1)}(5))^2\over1458} +{10\over 243} (161+432 \pi a^{(1)}(5)) \log\left(r^2C_E^2\over
                 G_Nm\right) \cr
                 &+ {100\over 81}\log\left(r^2C_E^2\over
                 G_Nm\right)^2
                 \Bigg)    \left(G_N m \over\pi r^2\right)^4 +O(d-4),\cr
 h^{\rm dD(4)}_{2}&=  \Bigg({-19048-141696 \pi  a^{(1)}(5)373248 (\pi a^{(1)}(5))^2\over 729} +{160\over 243} (-41+216 \pi a^{(1)}(5)) \log\left(r^2C_E^2\over
                 G_Nm\right) \cr
                 &+ {800\over 81}\log\left(r^2C_E^2\over
                 G_Nm\right)^2
                 \Bigg)    \left(G_N m \over\pi r^2\right)^4 +O(d-4)\,.
\end{align}

\subsection{The Schwarzschild-Tangherlini  metric in six dimensions}

The metric components in six dimensions ($d=5$) are finite. They are
given up to the order $O(G_N^4)$  
in~\eqref{e:h2finite6D} and are reproduced by the sum of  the
contributions of the
tree-level amplitude in~(\ref{e:htree}), one-loop amplitude 
in~(\ref{e:honeloop}), two-loop amplitude in~(\ref{e:htwoloopdiv}) and
three-loop amplitude in~(\ref{e:hthreeloopdiv}) and setting $d=5$ in
these expressions.

%%%%%%%%%%%%%%%%%%%%%%%%%%%%%%%%%%%%%%%%%%%%%%%%%%%%%%%%%%%%%%%%%%
\section{Discussion} \label{sec:discussion}

General relativity can be considered in  space-times of various
dimensions. It is therefore important to validate our current
understanding of the connection between scattering amplitudes and
classical general relativity in general
dimensions~\cite{KoemansCollado:2019ggb,Cristofoli:2020uzm} 

We have shown how to reconstruct the classical
Schwarzschild-Tangherlini  metric from scattering amplitudes in four,
five  and six dimensions. We have 
extracted the classical contribution as defined
in~\cite{Bjerrum-Bohr:2018xdl} from the vertex function for the
emission of a graviton from a massive scalar field. For such a static metric, the classical contribution is
obtained by taking appropriate residues on the time components of the
loop momenta. These residues project the quantum scattering amplitude
on contribution similar to the quantum tree graphs considered
in~\cite{Duff:1973zz}, by cutting the massive propagators. 

The amplitudes develop ultraviolet
divergences which are  renormalised by introducing higher-derivative non-minimal
couplings in~(\ref{e:Sctn}). The non-minimal coupling  removes the
ultraviolet divergences in the stress-tensor and the metric
components. For the static solution the higher $n\geq2$ non-minimal
coupling only contribute from insertions in tree-level graphs.
Interestingly, in six dimensions the metric components are finite but
the stress-tensor has ultraviolet divergences. These divergences are
removed by adding counter-terms from non-minimal couplings. These
counter-terms do not induce any contribution to the metric components.
From  the presence of ultraviolet poles in the master integrals
$J_{(l)}(\vec q^2)$ in~\eqref{e:Jnresult}, we conclude that in all dimensions one needs
to introduce an infinite set of higher-derivative non-minimal operators for
removing the ultraviolet divergences from the scattering
amplitude.
These  counter-terms do not affect the
space-time geometry because their effect  is reabsorbed by the change of coordinate from the
de Donder coordinate system to the Schwarzschild-Tangherlini coordinate
system.

The scattering amplitude approach presented in this work can be applied to
any effective field theory of gravity coupled to matter fields.
The amplitudes computations, being performed in general
dimensions, lead to results that have an analytic dependence on the
space-time dimensions.  As black-hole solutions develop non trivial properties in general
dimensions~\cite{Emparan:2008eg,Emparan:2013moa}, it is interesting to
apply the method of this paper to other black-hole metrics. The Kerr-Newman and Reissner-Nordstr\"om metric in four dimensions have
been obtained
in~\cite{Donoghue:2001qc,BjerrumBohr:2002ks,Guevara:2018wpp,Chung:2018kqs,Moynihan:2019bor,Chung:2019yfs,Guevara:2019fsj,Cristofoli:2020hnk} by 
considering  tree-level and one-loop vertex function of the
emission of the graviton from a massive particle of  spin $s$. The
higher order post-Minkowskian contributions should be obtained from
higher-loop amplitudes in a direct application of the methods used in
this work.

%%%%%%%%%%%%%%%%%%%%%%%%%%%%%%%%%%%%%%%%%%%%%%%%%%%%%%%%%%%%%%%%%%
\acknowledgments 

We would like to thank Emil Bjerrum-Bohr, Poul Damgaard, Paolo di Vecchia, Ludovic Plant\'e for discussions
and comments.  The research of P. Vanhove has received funding from
the ANR grant ``Amplitudes'' ANR-17- CE31-0001-01, and is partially
supported by Laboratory of Mirror Symmetry NRU HSE, RF Government
grant, ag. N$^\circ$ 14.641.31.0001. P.V. is grateful to the
I.H.E.S. for allowing to use their computer resources.

\appendix
\section{Fourier transforms }\label{sec:FT}

Here we collect the Fourier integrals used to calculate the long range corrections to
the energy momentum tensor and the metric.

The Fourier transform form momentum space to direct space
\begin{equation}\label{e:FTinv}
   {\mathcal F}(\alpha,d)= \int_{\mathbb R^d} {1\over |q|^\alpha} e^{i\vec q\cdot\vec x}
    {d^d\vec q\over (2\pi)^d}={1\over (4\pi)^{d\over2}}
    {\Gamma\left(d-\alpha\over2\right)\over
      \Gamma\left(\alpha\over2\right)} \left(2\over |\vec x|\right)^{d-\alpha}\,.
  \end{equation}
Using that
\begin{equation}
  \partial_{x^i}\partial_{x^j} (\vec x^2)^\alpha= 2\alpha (\vec
  x^2)^{\alpha-1}\left( \delta_{ij} +2(\alpha-1) {x_ix_j\over \vec x^2}\right)\,,
\end{equation}
we have that 
\begin{equation}\label{e:FTqq}
 \mathcal F_{ij}(\alpha,d):=\int_{\mathbb R^d} {q_iq_j\over |\vec q|^{\alpha+2}} e^{i\vec q\cdot\vec x}
 {d^d\vec q\over(2\pi)^d}=
 {\mathcal F}(\alpha,d) \left({1\over\alpha}\delta_{ij}+
   {\alpha-d\over\alpha} {x_ix_j\over \vec x^2}\right)\,.
\end{equation}

We have in particular that
\begin{equation}
      {\mathcal F}(0,d)=0, \qquad \mathcal F_{ij}(0,d)=  
{\Gamma\left(d\over2\right)\over 2\pi^{d\over2}\, |\vec x|^d}\left(\delta_{ij}
   -d {x_ix_j\over \vec x^2}\right)\,.
\end{equation}

%-------------------------------------------------------------------------
\section{Vertices and Propagators}\label{sec:vertices}
We will here list the Feynman rules which are employed in our
calculation. For
the derivation of these forms, see~\cite{BjerrumBohr:2002ks,DeWitt:1967yk,DeWitt:1967ub,DeWitt:1967uc,Bjerrum-Bohr:2013bxa,Donoghue:1995cz,Sannan:1986tz}.  Our
convention differs from these work by having all incoming momenta.
We have stripped off factors of $i\sqrt{8\pi G_N}$ from the
vertices and made them explicit in the amplitudes.
\begin{itemize}
\item The massive scalar propagator is $\displaystyle \frac i{q^2-m^2+i\varepsilon}\,.$
\item The graviton propagator in de Donder gauge can be written in the
  form $\displaystyle \frac {i{\mathcal
P}^{\alpha\beta,\gamma\delta}}{q^2+i\varepsilon}$
where  $\mathcal P^{\alpha\beta,\gamma\delta}$ is defined by
\end{itemize}
\begin{equation}\label{e:Pdef}
  \mathcal P^{\mu\nu,\rho\sigma}=\frac12\,
  \left(\eta^{\mu\rho}\eta^{\nu\sigma}+\eta^{\mu\sigma}\eta^{\nu\rho}-{2\over
      D-2}\eta^{\mu\nu}\eta^{\rho\sigma}\right)
\end{equation}
\begin{itemize}
\item  The 2-scalar-1-graviton vertex $ \tau_1^{\mu
    \nu}(p_1,p_2)$ is
\begin{equation}\label{e:tau1}
\tau^{\mu\nu}(p_1,p_2) =p_1^\mu p_2^{\nu}
+p_1^\nu p_2^{\mu} +\frac12 \eta^{\mu\nu}\,(p_1-p_2)^2\,.
\end{equation}
\item The three-graviton vertex has been derived in~\cite{Donoghue:1995cz}, where $k+q+\pi=0$,
\begin{equation}\label{e:tau3}{
\begin{aligned}
{\tau_{(3)}}_{\alpha  \beta ,\gamma \delta }^{\mu
\nu}(k,q)&=-
\bigg({\cal P}_{\alpha \beta \gamma \delta }\bigg[k^\mu k^\nu+ \pi^\mu\pi^\nu
+q^\mu q^\nu-
\frac32\eta^{\mu \nu}q^2\bigg]\\[0.00cm]&
+2q_\lambda q_\sigma\bigg[ I_{\alpha \beta }^{\ \ \
\sigma\lambda}I_{\gamma \delta
}^{\ \ \ \mu \nu} + I_{\gamma \delta }^{\ \ \ \sigma\lambda}I_{\alpha
\beta }^{\ \ \
\mu \nu} -I_{\alpha \beta }^{\ \ \ \mu  \sigma} I_{\gamma \delta }^{\ \ \
\nu
\lambda} - I_{\gamma \delta }^{\ \ \ \mu \sigma} I_{\alpha \beta }^{\ \ \
\nu
\lambda}
\bigg]\\[0cm]&
+\bigg[q_\lambda q^\mu \bigg(\eta_{\alpha \beta }I_{\gamma \delta }^{\ \ \
\nu
\lambda}+\eta_{\gamma \delta }I_{\alpha \beta }^{\ \ \ \nu
\lambda}\bigg) +q_\lambda
q^\nu \left(\eta_{\alpha \beta }I_{\gamma \delta }^{\ \ \ \mu
\lambda}+\eta_{\gamma
\delta }I_{\alpha \beta }^{\ \ \ \mu  \lambda}\right)\\&
-q^2\left(\eta_{\alpha
\beta }I_{\gamma \delta }^{\ \ \ \mu \nu}+\eta_{\gamma \delta }I_{\alpha
\beta }^{\
\ \ \mu \nu}\right) -\eta^{\mu \nu}q_\sigma q_\lambda\left(\eta_{\alpha
\beta
}I_{\gamma \delta }^{\ \ \ \sigma\lambda} +\eta_{\gamma \delta }I_{\alpha
\beta }^{\
\ \
\sigma\lambda}\right)\bigg]\\[0cm]&
+\bigg[2q_\lambda\big(I_{\alpha \beta }^{\ \ \ \lambda\sigma}I_{\gamma
\delta
\sigma}^{\ \ \ \ \nu}\pi^\mu +I_{\alpha \beta }^{\ \ \
\lambda\sigma}I_{\gamma
\delta \sigma}^{\ \ \ \ \mu }\pi^\nu +I_{\gamma \delta }^{\ \ \
\lambda\sigma}I_{\alpha \beta \sigma}^{\ \ \ \ \nu}k^\mu +I_{\gamma \delta
}^{\ \ \
\lambda\sigma}I_{\alpha \beta \sigma}^{\ \ \ \ \mu }k^\nu \big)\\&
+q^2\left(I_{\alpha \beta \sigma}^{\ \ \ \ \mu }I_{\gamma \delta }^{\ \ \
\nu
\sigma} + I_{\alpha \beta }^{\ \ \ \nu \sigma}I_{\gamma \delta \sigma}^{\
\ \ \ \mu
}\right) +\eta^{\mu \nu}q_\sigma q_\lambda\left(I_{\alpha \beta }^{\ \ \
\lambda\rho}I_{\gamma \delta  \rho}^{\ \ \ \ \sigma} +I_{\gamma \delta
}^{\ \ \
\lambda\rho}I_{\alpha \beta  \rho}^{\ \ \ \
\sigma}\right)\bigg]\\[0cm]&
+\bigg\{(k^2+\pi^2)\big[\mathcal P_{\alpha \beta }^{\ \ \ \mu  \sigma}\mathcal P_{\gamma
\delta,
\sigma}^{\ \ \ \ \nu} +\mathcal P_{\gamma \delta }^{\ \ \ \mu  \sigma}\mathcal P_{\alpha
\beta,
\sigma}^{\ \ \ \ \nu} -\frac12\eta^{\mu \nu}({\cal P}_{\alpha \beta ,\gamma
\delta
}-\eta_{\alpha\beta}\eta_{\gamma\delta})\big]\\&+\left(\mathcal P_{\gamma \delta }^{\ \ \ \mu \nu}\eta_{\alpha \beta
}\pi^2+\mathcal P_{\alpha
\beta }^{\ \ \ \mu \nu}\eta_{\gamma \delta }k^2\right)\bigg\}\bigg)\,,
\end{aligned}}
\end{equation}
where $  I_{\alpha\beta,\gamma\delta}:= \mathcal P_{\alpha\beta,\gamma\delta}+
  \frac12\, \eta_{\alpha\beta}\eta_{\gamma\delta}$.
These vertices are equivalent to the ones
computed with the vertices given by De Witt~\cite{DeWitt:1967yk,DeWitt:1967ub,DeWitt:1967uc} and
Sannan~\cite{Sannan:1986tz}.  We remark that the expression for
$\tau_3$ is simpler than the three-graviton vertex in these references.

We notice that the three-graviton vertex satisfies the identity 
\begin{equation}\label{e:tau3identity}
\tau^{\mu\nu}_{(3)\
    \pi\rho,\sigma\tau}(l,q)\mathcal{P}^{\pi\rho}_{\alpha\beta}\mathcal{P}^{\sigma\tau}_{\gamma\delta}= \tau^{\mu\nu}_{(3)\    \alpha\beta,\gamma\delta}(l,q)
\end{equation}
that will be used to simplify the expression of the amplitude.

\item The four-graviton vertex with  $k_1+k_2+k_3+k_4=0$ is given in~\cite{Sannan:1986tz,PlanteThesis}
\begin{equation}\label{e:tau4}{
\begin{aligned}
{\tilde \tau_{(4)\,\mu \nu ,\sigma \tau ,  \iota \kappa}}^{\rho
\lambda}(k_1,k_2,k_3,k_4)&= -\frac{1}{32} (k_1 \cdot k_2 \eta^{\mu \nu} \eta^{\sigma \tau}  \eta^{\rho  \lambda} \eta_{\iota  \kappa} ) - \frac{1}{16} (k_1^{\sigma} k_1^{\tau} \eta^{\mu \nu} \eta^{\rho  \lambda} \eta_{\iota  \kappa} ) \\[0.00cm]&
- \frac{1}{16} (k_1^{\sigma} k_2^{\mu} \eta^{\nu \tau}\eta^{\rho  \lambda} \eta_{\iota  \kappa} ) + \frac{1}{32} (k_1 \cdot k_2 \eta^{\mu \sigma} \eta^{\nu \tau} \eta^{\rho  \lambda} \eta_{\iota  \kappa} ) 
 + \frac{1}{16} (k_1 \cdot k_2 \eta^{\mu \nu}  \eta^{\sigma \tau}  \eta^\rho_\iota\eta^\lambda_\kappa) \\[0.00cm]&+ \frac{1}{8} (k_1^{\sigma} k_1^{\tau} \eta^{\mu \nu}  \eta^\rho_\iota \eta^\lambda_\kappa) 
 + \frac{1}{8} (k_1^{\sigma} k_2^{\mu} \eta^{\nu \tau} \eta^\rho_\iota \eta^\lambda_\kappa) - \frac{1}{16} (k_1 \cdot k_2 \eta^{\mu \sigma} \eta^{\nu \tau} \eta^\rho_\iota \eta^\lambda_\kappa) \\[0.00cm]&
 + \frac{1}{4} (k_1 \cdot k_2 \eta^{\mu \nu}  \eta^{\sigma \rho} \eta{\tau \lambda} \eta_{\iota  \kappa} ) + \frac{1}{4} (k_1^{\sigma} k_1^{\tau} \eta^{\mu \rho} \eta^{\nu \lambda} \eta_{\iota  \kappa} ) 
 + \frac{1}{8} (k_1^{\rho} k_2^{\lambda} \eta^{\mu \sigma} \eta^{\nu \tau} \eta_{\iota  \kappa} )\\[0.00cm]& + \frac{1}{2} (k_1^{\sigma} k_2^{\rho} \eta^{\tau \mu} \eta^{\nu \lambda} \eta_{\iota  \kappa} ) 
 - \frac{1}{4} (k_1 \cdot k_2 \eta^{\nu \sigma} \eta^{\tau \rho} \eta^{\lambda \mu} \eta_{\iota  \kappa} ) + \frac{1}{4} (k_1^{\sigma} k_2^{\mu} \eta^{\tau \rho} \eta^{\lambda \nu} \eta_{\iota  \kappa} ) \\[0.00cm]&
+ \frac{1}{4} (k_1^{\sigma} k_1^{\rho} \eta^{\tau \lambda} \eta^{\mu \nu} \eta_{\iota  \kappa} ) - \frac{1}{2} (k_1 \cdot k_2 \eta^{\mu \nu} \eta^{\tau \rho} \eta^\lambda_\iota \eta_\kappa^\sigma)
- \frac{1}{2} (k_1^{\sigma} k_1^{\tau} \eta^{\nu \rho} \eta^\lambda_\iota \eta_\kappa^\mu) \\[0.00cm]& -\frac{1}{2} (k_1^{\rho} k_2^{\lambda} \eta^\nu_\iota \eta_\kappa^\sigma \eta^{\tau \mu})
 - (k_1^{\sigma} k_2^{\rho} \eta^\tau_\iota \eta_\kappa^\mu \eta^{\nu \lambda}) - \frac{1}{2} (k_1^{\rho} k_{2\,\iota} \eta^{\lambda \sigma} \eta^{\tau \mu} \eta^\nu_\kappa) \\[0.00cm]&
+ \frac{1}{4}(k_1 \cdot k_2 \eta^{\nu \rho} \eta^{\lambda \sigma} \eta^\tau_\iota \eta_\kappa^\mu) - \frac{1}{2} (k_1^{\sigma} k_1^{\rho} \eta^{\mu \nu}  \eta^\tau_\iota \eta_\kappa^\lambda)
- \frac{1}{4} (k_1 \cdot k_2 \eta^{\mu \rho} \eta^{\nu \lambda} \eta^\sigma_\iota \eta^\tau_\kappa) \\[0.00cm]&- \frac{1}{2} (k_1^{\sigma} k_1^{\rho} \eta^{\tau \lambda} \eta^\mu_\iota \eta^\nu_\kappa) 
- \frac{1}{4} (k_1^{\rho} k_{2\, \iota} \eta^\lambda_\kappa \eta^{\mu \sigma} \eta^{\nu \tau}) - (k_1^{\sigma} k_2^{\rho} \eta^{\tau \mu} \eta^\nu_\iota \eta_\kappa^\lambda) \\[0.00cm]&
-\frac{1}{2} (k_1^{\sigma} k_2^{\mu} \eta^{\tau \rho} \eta^\lambda_\iota \eta_\kappa^\nu) + \frac{1}{2}(k_1 \cdot k_2 \eta^{\nu \sigma} \eta^{\tau \rho} \eta^\lambda_\iota \eta_\kappa^\mu) 
+ Sym(k_1,k_2,k_3,k_4)
\end{aligned}}
\end{equation}

we introduce  the short hand notation 
\begin{equation}
  \tau^{\mu\nu}_{(4)\, \gamma\delta,\sigma\tau,\iota\theta}(k_1,k_2,k_3,k_4):=\tilde
  \tau^{\mu\nu}_{(4)\, \alpha\beta, \gamma\delta, \epsilon\eta}(k_1,k_2,k_3,k_4) \mathcal P^{\alpha\beta}_{ \gamma\delta}
  \mathcal P^{\gamma\delta}_{\sigma\tau} \mathcal P^{\epsilon\eta}_{\iota\theta}\,.
\end{equation}

\end{itemize}
%%%%%%%%%%%%%%%%%%%%%%%%%%%%%%%%%%%%%%%%%%%%%%%%%%%%%%%%%%%%%%%%%


\begin{thebibliography}{99}
 
\bibitem{tHooft:1974toh}
G.~'t Hooft and M.~J.~G.~Veltman,
``One Loop Divergencies in the Theory of Gravitation,''
Ann. Inst. H. Poincare Phys. Theor. A \textbf{20} (1974), 69-94


\bibitem{Veltman:1975vx}
  M.~J.~G.~Veltman,
  ``Quantum Theory of Gravitation,''
  Conf.\ Proc.\ C {\bf 7507281} (1975) 265.
  %%CITATION = CONFP,C7507281,265;%%

  
%\cite{DeWitt:1967yk}
\bibitem{DeWitt:1967yk}
B.~S.~DeWitt,
``Quantum Theory of Gravity. 1. The Canonical Theory,''
Phys. Rev. \textbf{160} (1967), 1113-1148


%\cite{DeWitt:1967ub}
\bibitem{DeWitt:1967ub}
B.~S.~DeWitt,
``Quantum Theory of Gravity. 2. The Manifestly Covariant Theory,''
Phys. Rev. \textbf{162} (1967), 1195-1239

%\cite{DeWitt:1967uc}
\bibitem{DeWitt:1967uc}
B.~S.~DeWitt,
``Quantum Theory of Gravity. 3. Applications of the Covariant Theory,''
Phys. Rev. \textbf{162} (1967), 1239-1256

  
  
\bibitem{Donoghue:1994dn}
J.~F.~Donoghue,
``General Relativity as an Effective Field Theory: the Leading Quantum Corrections,''
Phys. Rev. D \textbf{50} (1994), 3874-3888
[arXiv:gr-qc/9405057 [gr-qc]].


  \bibitem{Duff:1973zz}
M.~J.~Duff, ``Quantum Tree Graphs and the Schwarzschild Solution,''
Phys. Rev. D \textbf{7} (1973), 2317-2326


\bibitem{Iwasaki:1971vb}
Y.~Iwasaki,
``Quantum Theory of Gravitation Vs. Classical Theory. - Fourth-Order Potential,''
Prog. Theor. Phys. \textbf{46} (1971), 1587-1609


%
\bibitem{BjerrumBohr:2002ks}
N.~E.~J.~Bjerrum-Bohr, J.~F.~Donoghue and B.~R.~Holstein,
``Quantum corrections to the Schwarzschild and Kerr metrics,''
Phys. Rev. D \textbf{68} (2003), 084005
[arXiv:hep-th/0211071 [hep-th]].



\bibitem{Holstein:2004dn}
  B.~R.~Holstein and J.~F.~Donoghue,
  ``Classical Physics and Quantum Loops,''
  Phys.\ Rev.\ Lett.\  {\bf 93} (2004) 201602
  [hep-th/0405239].
  %% CITATION = doi:10.1103/PhysRevLett.93.201602;%%

\bibitem{Donoghue:1996mt}
J.~F.~Donoghue and T.~Torma,
``On the Power Counting of Loop Diagrams in General Relativity,''
Phys. Rev. D \textbf{54} (1996), 4963-4972
[arXiv:hep-th/9602121 [hep-th]].

  \bibitem{Bjerrum-Bohr:2018xdl}
  N.~E.~J.~Bjerrum-Bohr, P.~H.~Damgaard, G.~Festuccia, L.~Plant\'e and P.~Vanhove,
  ``General Relativity from Scattering Amplitudes,''
  Phys.\ Rev.\ Lett.\  {\bf 121} (2018) no.17,  171601
  [arXiv:1806.04920 [hep-th]].
  %%CITATION = doi:10.1103/PhysRevLett.121.171601;%%
  
\bibitem{Kosower:2018adc}
D.~A.~Kosower, B.~Maybee and D.~O'Connell,
``Amplitudes, Observables, and Classical Scattering,''
JHEP \textbf{02} (2019), 137
%doi:10.1007/JHEP02(2019)137
[arXiv:1811.10950 [hep-th]].


\bibitem{Cheung:2018wkq}
C.~Cheung, I.~Z.~Rothstein and M.~P.~Solon,
``From Scattering Amplitudes to Classical Potentials in the Post-Minkowskian Expansion,''
Phys. Rev. Lett. \textbf{121} (2018) no.25, 251101
[arXiv:1808.02489 [hep-th]].

 
\bibitem{Bern:2019nnu}
Z.~Bern, C.~Cheung, R.~Roiban, C.~H.~Shen, M.~P.~Solon and M.~Zeng,
``Scattering Amplitudes and the Conservative Hamiltonian for Binary Systems at Third Post-Minkowskian Order,''
Phys. Rev. Lett. \textbf{122} (2019) no.20, 201603
[arXiv:1901.04424 [hep-th]].

\bibitem{Bern:2019crd}
Z.~Bern, C.~Cheung, R.~Roiban, C.~H.~Shen, M.~P.~Solon and M.~Zeng,
``Black Hole Binary Dynamics from the Double Copy and Effective Theory,''
JHEP \textbf{10} (2019), 206
[arXiv:1908.01493 [hep-th]].


\bibitem{Chung:2019duq}
M.~Z.~Chung, Y.~T.~Huang and J.~W.~Kim,
``Classical Potential for General Spinning Bodies,''
JHEP \textbf{09} (2020), 074
[arXiv:1908.08463 [hep-th]].


\bibitem{Kalin:2019rwq}
G.~K\"alin and R.~A.~Porto,
``From Boundary Data to Bound States,''
JHEP \textbf{01} (2020), 072
[arXiv:1910.03008 [hep-th]].

\bibitem{Cheung:2020gyp}
C.~Cheung and M.~P.~Solon,
``Classical Gravitational Scattering at $ \mathcal{O} (G^{3})$ from Feynman Diagrams,''
JHEP \textbf{06} (2020), 144
[arXiv:2003.08351 [hep-th]].

\bibitem{1821624}
G.~Mogull, J.~Plefka and J.~Steinhoff,
``Classical Black Hole Scattering from a Worldline Quantum Field Theory,''
[arXiv:2010.02865 [hep-th]].


\bibitem{Collado:2018isu}
A.~K.~Collado, P.~Di Vecchia, R.~Russo and S.~Thomas,
``The Subleading Eikonal in Supergravity Theories,''
JHEP \textbf{10} (2018), 038
[arXiv:1807.04588 [hep-th]].

\bibitem{KoemansCollado:2019ggb}
A.~Koemans Collado, P.~Di Vecchia and R.~Russo,
``Revisiting the Second Post-Minkowskian Eikonal and the Dynamics of Binary Black Holes,''
Phys. Rev. D \textbf{100} (2019) no.6, 066028
[arXiv:1904.02667 [hep-th]].

\bibitem{Cristofoli:2020uzm}
A.~Cristofoli, P.~H.~Damgaard, P.~Di Vecchia and C.~Heissenberg,
``Second-Order Post-Minkowskian Scattering in Arbitrary Dimensions,''
JHEP \textbf{07} (2020), 122
[arXiv:2003.10274 [hep-th]].
  
 \bibitem{Jakobsen:2020ksu}
G.~U.~Jakobsen,
``Schwarzschild-Tangherlini Metric from Scattering Amplitudes,''
[arXiv:2006.01734 [hep-th]].


\bibitem{Emparan:2008eg}
R.~Emparan and H.~S.~Reall,
``Black Holes in Higher Dimensions,''
Living Rev. Rel. \textbf{11} (2008), 6
[arXiv:0801.3471 [hep-th]].

\bibitem{Emparan:2013moa}
R.~Emparan, R.~Suzuki and K.~Tanabe,
``The Large D Limit of General Relativity,''
JHEP \textbf{06} (2013), 009
[arXiv:1302.6382 [hep-th]].

  \bibitem{Goldberger:2004jt}
W.~D.~Goldberger and I.~Z.~Rothstein,
``An Effective Field Theory of Gravity for Extended Objects,''
Phys. Rev. D \textbf{73} (2006), 104029
%doi:10.1103/PhysRevD.73.104029
[arXiv:hep-th/0409156 [hep-th]].

\bibitem{BjerrumBohr:2006mz}
N.~E.~J.~Bjerrum-Bohr, J.~F.~Donoghue and B.~R.~Holstein,
``On the Parameterization Dependence of the Energy Momentum Tensor and the Metric,''
Phys. Rev. D \textbf{75} (2007), 108502
[arXiv:gr-qc/0610096 [gr-qc]].

  
\bibitem{Fromholz:2013hka}
P.~Fromholz, E.~Poisson and C.~M.~Will,
``The Schwarzschild Metric: It's the Coordinates, Stupid!,''
Am. J. Phys. \textbf{82} (2014), 295
[arXiv:1308.0394 [gr-qc]].

\bibitem{nobelpenrose} Citation for the 2020 Nobel prize award to 
  Roger Penrose \url{https://www.nobelprize.org/prizes/physics/2020/press-release/}

\bibitem{PlanteThesis} L.~Plant\'e, ``Some aspects on effective field
  theory of gravity'',  Phd thesis, Universit\'e Paris-Saclay 2016


\bibitem{Galusha} J.~Galusha, ``Investigations into the Classical Limit of
  Quantum Field Theory'',   Master thesis Niels Bohr International
  Academy 2018
  
\bibitem{Levy:1969cr}
M.~Levy and J.~Sucher,
``Eikonal Approximation in Quantum Field Theory,''
Phys. Rev. \textbf{186} (1969), 1656-1670

\bibitem{Lee:2012cn}
R.~N.~Lee,
``Presenting Litered: a Tool for the Loop Integrals Reduction,''
[arXiv:1212.2685 [hep-ph]].


\bibitem{Lee:2013mka}
R.~N.~Lee,
``Litered 1.4: a Powerful Tool for Reduction of Multiloop Integrals,''
J. Phys. Conf. Ser. \textbf{523} (2014), 012059
[arXiv:1310.1145 [hep-ph]].


  \bibitem{Vanhove:2014wqa}
P.~Vanhove,
``The Physics and the Mixed Hodge Structure of Feynman Integrals,''
Proc. Symp. Pure Math. \textbf{88} (2014), 161-194
[arXiv:1401.6438 [hep-th]].
  


\bibitem{Cristofoli:2019neg}
A.~Cristofoli, N.~E.~J.~Bjerrum-Bohr, P.~H.~Damgaard and P.~Vanhove,
``Post-Minkowskian Hamiltonians in General Relativity,''
Phys. Rev. D \textbf{100} (2019) no.8, 084040
[arXiv:1906.01579 [hep-th]].

\bibitem{Lagarias} J.C. Lagarias. ``Euler's constant: Euler's work and modern developments.'' Bull. Amer. Math. Soc. (N.S.)
 {\bf 50} (2013), no. 4, 527–628.

  \bibitem{Prinz:2020nru}
D.~Prinz,
``Gravity-Matter Feynman Rules for Any Valence,''
[arXiv:2004.09543 [hep-th]].

\bibitem{Tangherlini:1963bw}
F.~R.~Tangherlini,
``Schwarzschild Field in $N$ Dimensions and the Dimensionality of Space Problem,''
Nuovo Cim. \textbf{27} (1963), 636-651

\bibitem{Donoghue:2001qc}
J.~F.~Donoghue, B.~R.~Holstein, B.~Garbrecht and T.~Konstandin,
``Quantum Corrections to the Reissner-Nordstrom and Kerr-Newman Metrics,''
Phys. Lett. B \textbf{529} (2002), 132-142
[erratum: Phys. Lett. B \textbf{612} (2005), 311-312]
[arXiv:hep-th/0112237 [hep-th]].

\bibitem{Guevara:2018wpp}
A.~Guevara, A.~Ochirov and J.~Vines,
``Scattering of Spinning Black Holes from Exponentiated Soft Factors,''
JHEP \textbf{09} (2019), 056
[arXiv:1812.06895 [hep-th]].

\bibitem{Chung:2018kqs}
M.~Z.~Chung, Y.~T.~Huang, J.~W.~Kim and S.~Lee,
``The Simplest Massive S-Matrix: from Minimal Coupling to Black Holes,''
JHEP \textbf{04} (2019), 156
[arXiv:1812.08752 [hep-th]].

 \bibitem{Moynihan:2019bor}
N.~Moynihan,
``Kerr-Newman from Minimal Coupling,''
JHEP \textbf{01} (2020), 014
[arXiv:1909.05217 [hep-th]].

\bibitem{Chung:2019yfs}
M.~Z.~Chung, Y.~T.~Huang and J.~W.~Kim,
``Kerr-Newman Stress-Tensor from Minimal Coupling to All Orders in Spin,''
[arXiv:1911.12775 [hep-th]].
 
\bibitem{Guevara:2019fsj}
A.~Guevara, A.~Ochirov and J.~Vines,
``Black-Hole Scattering with General Spin Directions from Minimal-Coupling Amplitudes,''
Phys. Rev. D \textbf{100} (2019) no.10, 104024
[arXiv:1906.10071 [hep-th]].

\bibitem{Cristofoli:2020hnk}
A.~Cristofoli,
``Gravitational Shock Waves and Scattering Amplitudes,''
[arXiv:2006.08283 [hep-th]].
  
  \bibitem{Bjerrum-Bohr:2013bxa}
N.~E.~J.~Bjerrum-Bohr, J.~F.~Donoghue and P.~Vanhove,
``On-Shell Techniques and Universal Results in Quantum Gravity,''
JHEP \textbf{02} (2014), 111
[arXiv:1309.0804 [hep-th]].


\bibitem{Donoghue:1995cz}
J.~F.~Donoghue,
``Introduction to the Effective Field Theory Description of Gravity,''  in \emph{{Advanced School on Effective Theories Almunecar,
  Spain, June 25-July 1, 1995}}, 1995.
[arXiv:gr-qc/9512024 [gr-qc]].


%\cite{Sannan:1986tz}
\bibitem{Sannan:1986tz}
S.~Sannan,
``Gravity as the Limit of the Type {II} Superstring Theory,''
Phys. Rev. D \textbf{34} (1986), 1749


\end{thebibliography}
\end{document}